\documentclass[pra, superscriptaddress, twocolumn, showpacs, floatfix, nofootinbib]{revtex4}

\usepackage{amsfonts}
\usepackage{amssymb}
\usepackage{graphicx}
\usepackage{color}
\usepackage{amsmath}
\usepackage[bookmarksnumbered, bookmarks, breaklinks, linktocpage]{hyperref}

\newcommand{\Tr}        {\mathrm{Tr}}
\newcommand{\Id}        {I}
\newcommand{\bra}[1]    {\langle #1|}
\newcommand{\ket}[1]    {| #1 \rangle}
\newcommand{\bk}[2]     {\langle #1 | #2 \rangle}
\newcommand{\kb}[2]     {| #1 \rangle \! \langle #2 |}
\newcommand{\cH}        {{\mathcal H}}
\newcommand{\cS}        {{\mathcal S}}
\newcommand{\cA}        {{\mathcal A}}
\newcommand{\cE}        {{\mathcal E}}

\newcommand{\eend}      {\hspace{\stretch{1}}\rule{1ex}{1ex}}
\def\bA{{\bf A}}
\def\bB{{\bf B}}
\def\bC{{\bf C}}
\def\bX{{\bf X}}
\def\bY{{\bf Y}}
\def\bZ{{\bf Z}}
\def\bH{{\bf H}}
\newcommand\cF{{\mathcal F}}

\newcommand\dpcom[1]{}

\newtheorem{definition}{Definition}[section]

\newtheorem{lemma}{Lemma}[section]
\newtheorem{theorem}{Theorem}[section]
\newenvironment{proof}{\noindent {\bf Proof} }{\eend}
\newtheorem{proposition}{Proposition}[section]

\newtheorem{corollary}{Corollary}[section]

\begin{document}

\title{Environment as a Witness: Selective Proliferation of Information and Emergence of
 Objectivity in a Quantum Universe}

\author{Harold Ollivier}
\affiliation{Perimeter Institute, 31 Caroline St N, Waterloo, ON N2L 2Y5, Canada.}
\affiliation{INRIA - Projet Codes, BP 105, F-78153 Le Chesnay, France}
\author{David Poulin\footnote{Present address: {\em School of Physical Sciences, The University of Queensland, Australia.}}}
\affiliation{Perimeter Institute, 31 Caroline St N, Waterloo, ON N2L 2Y5, Canada.}
\affiliation{Institute for Quantum Computing, University of Waterloo, ON Canada, N2L 3G1}
\author{Wojciech H. Zurek}
\affiliation{Theory Division, LANL, MS-B213, Los Alamos, NM  87545, USA}

\date{\today}

\begin{abstract}
We study the role of the information deposited in the environment of
an open quantum system in course of the decoherence process. Redundant
spreading of information --- the fact that some observables of the
system can be independently ``read-off'' from many distinct fragments
of the environment --- is investigated as the key to effective
objectivity, the essential ingredient of ``classical reality''. This
focus on the environment as a communication channel through which
observers learn about physical systems underscores importance of {\it
quantum Darwinism} --- selective proliferation of information about
``the fittest states'' chosen by the dynamics of decoherence at the
expense of their superpositions --- as redundancy imposes the
existence of preferred observables. We demonstrate that the only
observables that can leave multiple imprints in the environment are
the familiar {\it pointer observables} singled out by {\it
environment-induced superselection (einselection)} for their
predictability. Many independent observers monitoring the environment
will therefore agree on properties of the system as they can only
learn about preferred observables. In this operational sense, the
selective spreading of information leads to appearance of an objective
``classical reality'' from within quantum substrate.
\end{abstract}

\maketitle

\section{Introduction}
Emergence of a classical reality within the quantum Universe has been
the focus of discussions on the interpretation of quantum theory ever
since its inception.  Measurement --- the process through which we learn
about the world --- has the power to transform fuzzy quantum states into 
solid classical facts. Understanding measurements has been therefore rightly
regarded as the key to unlocking the mystery of the quantum-classical
transition since the early days \cite{WZ83a}. Bohr's interpretation proposed in
1928~\cite{Boh28a} introduced the classical domain ``by hand'', with a
demand that much of the Universe --- including measuring devices ---
must be classical. This {\it Copenhagen interpretation} proved to be
workable and durable, but is ultimately unsatisfying, because of the
arbitrary split between ``the quantum'' and ``the classical''.  Thus,
Copenhagen interpretation notwithstanding, attempts to explain the
emergence of the classical, objective reality (including measurement
outcomes) using only quantum theory were made ever since its structure
became clear in the late 1920's.

Von Neumann \cite{vN55a} introduced a particularly influential model
of the measurement process. In his approach --- and in contrast to
Bohr's view --- the apparatus $\cA$ is also quantum. ``Bit-by-bit
measurement'' \cite{Zur81a} is the simplest example: a 2-dimensional
system $\cS$ in pure state $\alpha \ket 0 + \beta \ket 1$ interacts
with a 2-dimensional apparatus initially in state $\ket{\mu_0^\cA}$.
In course of the controlled-not (or ``measurement'') gate the
apparatus becomes --- as one would now say --- entangled with the
system $\cS$:
\begin{equation}
(\alpha\ket0 + \beta\ket 1)\otimes \ket{\mu_0^\cA} \rightarrow
\alpha\ket0 \otimes\ket{\mu_0^\cA} + \beta\ket1\otimes\ket{\mu_1^\cA}.
\label{eq:ex_decoherence}
\end{equation} 
This is pre-measurement. It implies correlation of $\cS$ and $\cA$,
but does not yield a definite outcome. 

The structure of Eq.~(\ref{eq:ex_decoherence})
suggests a relative state interpretation of quantum theory
\cite{Eve57a}. However, to make contact with the familiar reality, one
must point out the ``preferred relative states''. Yet, in
the bipartite setting of the pre-measurement, such proposals are
difficult to make without some {\it ad hoc} assumptions (e.g, about
the special role of either ``memory states''~\cite{Eve57a} or of the
``Schmidt basis''~\cite{Zeh73a}).

Presence of entanglement in the state obtained after the
pre-measurement implies a {\em basis ambiguity} ---~correlation of
observables of $\cA$ with incompatible sets of pure states of the
system which cannot be resolved without some modification of the
model~\cite{ Zur81a}. For example, the states $\{\ket{\mu_0^\cA},\ket{\mu_1^\cA}\}$
of $\cA$ are in one-to-one correspondence with the states $\{\ket0 , \ket
1\}$ of $\cS$, while the states $\left\{(\ket{\mu_0^\cA} \pm
\ket{\mu_1^\cA})/\sqrt2\right\}$ are in one-to-one correspondence with
the states $\left\{(\alpha\ket 0 \pm \beta\ket 1)/\sqrt2\right\}$ of
$\cS$. Thus, von Neumann's model does not account for the existence of a fixed ``menu'' of
possible measurement outcomes --- an issue that must be addressed
before the apparent selection of one position on this menu (i.e.\@~the
``collapse of the wave-packet'') is contemplated.

Decoherence theory (see e.g.  ~\cite{Zur91a, GJK+96a, PZ01a, Zur03a,
Sch03a} for reviews) added a new element that goes well beyond the von
Neumann's measurement model: in addition to $\cS$ and $\cA$,
decoherence recognizes the role of the environment $\cE$ that
surrounds $\cA$ and interacts with the apparatus (or with any other
object immersed in ${\cE}$). The resulting ``openness'' of $\cA$
invalidates the egalitarian principle of superposition: while all
states in the Hilbert space of the apparatus $\cA$ are ``legal''
quantum superpositions, only some of them will retain their identity
--- will be stable in spite of the coupling to $\cE$.

Returning to
our example, the environment may interact with $\cA$ in such a way
that an arbitrary superposition $\xi_0 \ket{\mu_0^\cA} + \xi_1
\ket{\mu_1^\cA}$ is transformed into a mixture $|\xi_0|^2
\kb{\mu_0^\cA}{\mu_0^\cA} + |\xi_1|^2 \kb{\mu_1^\cA}{\mu_1^\cA}$ after
a very short time. Thus, only the two states $\ket{\mu_0^\cA}$ and
$\ket{\mu_1^\cA}$ remain pure over time. Selection of such preferred
set of states is known as environment-induced superselection, or {\em
einselection}. The persistence of the correlation between $\cA$ and
$\cS$ is the desired prerequisite of measurements, and only stable
{\em pointer basis} of $\cA$ selected by the interaction with $\cE$
fits the bill \cite{Zur81a, Zur82a, PZ01a, Zur03a, Sch03a}.  Indeed,
after the decoherence time, the joint state of $\cS$ and $\cA$,
Eq.~(\ref{eq:ex_decoherence}), becomes mixed:
\begin{equation}
|\alpha|^2 \kb 00 \otimes \kb{\mu_0^\cA}{\mu_0^\cA} + |\beta|^2 \kb 11
 \otimes \kb{\mu_1^\cA}{\mu_1^\cA}.
\end{equation}
As a consequence of decoherence, only classical correlations of $\cA$ with the system states
$\{\ket0,\ket1\}$ persist.

Understanding the reason for the loss of validity of the quantum
principle of superposition is a significant step in the understanding 
of the quantum-classical transition, but it does not go all the way in
justifying objectivity: the einselected pointer states are still
ultimately quantum. Thus, they remain sensitive to direct measurements
--- a purely quantum problem. An observer trying to find out about the
system directly will generally disturb its state, unless he happens to
make a non-demolition measurement \cite{CTD+80a,BK96a} in the pointer
basis.  As a consequence, it is effectively impossible for an
initially ignorant observer --- someone who does not know what are the
pointer states of the system --- to find out the 
state of a physical system through a direct measurement without
perturbing it: immediately after a direct measurement the state will be
what the observer finds out it is, but not --- in general --- what it
was before.

The situation becomes even more worrisome when one considers {\em
many} initially ignorant observers attempting to find out about the
system. As a consequence of the disturbance caused by a direct
measurement on the system, the information gained by the first
observer's measurement can get invalidated by the second observer's
measurement, etc., unless they all happen to measure commuting
observables --- or more precisely, unless the measured observables
share the system's pointer states as eigenstates.

Quantum subjectivity is to be contrasted with the objectivity of
classical physics, where many ignorant observers can --- at least in
principle --- find out the state of the system without modifying
it. This is because classical systems admit an underlying objective
description (``classical reality''), and classical states can be found
out by initially ignorant observers. This is generally not the case
for quantum systems. Thus, objective information about quantum systems
can be acquired directly by many only by a constrained set of
pre-agreed measurements on $\cS$ (see e.g.  \cite{Gra02a,Pou05a}).

Of course --- as noted in past discussions of einselection
\cite{Zur91a, Zur93b, Zur98a} --- there are good reasons for the
observers to focus on the set of states singled out by decoherence:
only such pointer states of $\cS$ (or their dynamically evolved
descendants) continue to faithfully describe the system in spite of
its interaction with $\cE$. All other states are affected by $\cE$,
making loss of predictability inevitable. Predictability is
characteristic of the states of classical systems, and is thus a
symptom of a classical reality. But, above all, prediction is the
reason for measurements.  One can therefore understand how observers
with practical experience with the emergent classicality (imposed in
our Universe by einselection) will be forced to choose the same
pointer observables as they make their choice of what to measure: save
for pointer observables, there is no other choice if measurements are
to be useful for prediction. This may look like a ``pre-agreement", but
it involves no consultation between observers: competing with the environment 
is simply not an option. 

In effect, the environment acts as a superobserver,
monitoring the same pointer observable over and over, with frequency
and accuracy that cannot be matched by other (e.g., human)
observers. They all have to measure observables that commute with the
pointer observable. Last but not least, interactions available to
observers are similar in structure (e.g., depend on distance, etc.) to
the interactions responsible for the einselection \cite{Zur98a,
Zur03a}. So, predictive utility in presence of decoherence and limited choice
of the Hamiltonians available in our Universe motivate ``pre-agreement" by 
constraining measurements to pointer observables.  However, even if such 
``pre-agreement imposed by einselection" can help single out what observables 
can become objective, the actual role of the environment in what happens in
practice far more dramatic and decisive: the environment is not just a
superobserver --- it becomes a witness. Observers use it to find out about 
systems of interest.  Hence, they must be content with the information 
that can be extracted from its fragments (as, generally, they will never 
be able to intercept all of $\cE$).

In its original formulation, decoherence theory treats the information
transferred to $\cE$ as inaccessible. However, in the real world, this
is typically {\em not} the case. Indeed, as was pointed out by one of
us some time ago \cite{Zur93b, Zur98a, Zur00a}, the fact that we gain
most of our information by intercepting a small fraction of the photon
environment is significant for the emergence of effectively classical
states from the quantum substrate. The purpose of this paper is to
investigate the consequences of such an indirect information
acquisition for the quantum-classical transition, and to explore the
relation of this ``environment as a witness'' \cite{Zur03a} point of
view to the predictability of the pointer states as well as to other
issues raised and partially explored in \cite{OPZ04b}. We shall
demonstrate that the manner in which the information is stored in the
environment is the reason for the inevitable consensus among many
observers about the state of the effectively classical (but ultimately
quantum) systems. In other words, the structure of information
deposition in {$\cE$} is responsible for the emergence of the
objective classical reality from the quantum substrate.   

We shall also begin to explore {\it quantum Darwinism}: the dynamics
responsible for the proliferation of correlations that leads to the
{\em survival of the fittest information}. This is a natural
complement to the environment as a witness approach that is focused on
how the data about {$\cS$} can be extracted by interrogating
{$\cE$}. Quantum Darwinism allows the environment to act as a witness
\cite{Zur03a, OPZ04b, Zur03b}, adding a new dimension to the modern
decoherence-based view of the emergence of the classical.

In the next section, we propose an operational notion of objectivity
and discuss how we will use it to investigate the quantum-classical
transition. Sections~\ref{sec:defs} and \ref{sec:info} set up
the notation and introduce tools of information theory required for
the present study. Section~\ref{sec:theory} contains the core
information-theoretic analysis of the manuscript. There, we establish
a number of facts about the structure of the information stored in the
environment, and study consequences of {\em redundant} imprinting of
selected system observables in $\cE$. These general properties are
then illustrated in Section~\ref{sec:model} on a dynamical model used
extensively in the study of decoherence. This allows us to extend the
results of our analysis, and establish a direct connection between
einselection and the emergence of an objective classical reality.
Finally, we consider some open questions in
Section~\ref{sec:discuss_model} and conclude in Section
\ref{sec:conclusion} with a summary.

\section{Operational definition of objectivity} 
\label{sec:motivation}

Quantum Darwinism \cite{Zur03a, OPZ04b} aims to show that a consensus
about the properties of a quantum system --- the key symptom of
classical reality --- arises naturally and inevitably from within
quantum theory when one recognizes the role of the environment as a
broadcasting medium that acquires --- in the process of monitoring the
system of interest that leads to decoherence and einselection ---
multiple copies of the information about preferred properties of the
system of interest.

We will set up a rigorous operational framework for the analysis of
the emergence of objective classical reality of quantum systems, based
on the following definition of objectivity:
\begin{definition}[Objective property]
A property of a physical system is {\em objective} when it is
\begin{enumerate}
\item simultaneously accessible to many observers,
\item who are able to find out what it is without prior knowledge
about the system of interest, and
\item who can arrive at a consensus about it without prior agreement.
\end{enumerate}
\end{definition}
This operational definition of what is objective is inspired by the
notion of ``element of physical reality'' used by Einstein, Podolsky
and Rosen in their famous ``EPR'' paper~\cite{EPR35a} on entanglement:
``{\em If, without in any way disturbing a system, we can predict with
certainty (i.e. with probability equal to unity) the value of a
physical quantity, then there exists an element of physical reality
corresponding to this physical quantity. [...] Regarded not as a
necessary, but merely as a sufficient condition of reality, this
criterion is in agreement with classical as well as quantum-mechanical
ideas of reality.}'' Any property of the system fulfilling this
requirement would be considered an element of objective reality
according to the definition we adopt.

The environment as a witness approach recognizes that in our everyday
acquisition of data, the decision of what to measure on $\cS$ is taken
out of our hands by the environment, and rests with the dynamics
responsible for decoherence --- i.e. for the monitoring of the system
by $\cE$. Nevertheless, in the absence of any structure of $\cE$, this
would still be insufficient to explain the emergence of a consensus
about the state of $\cS$.  Measurements on the environment suffer from
the same basis ambiguity problems as direct measurements on the
system: they can be performed in arbitrary bases. Moreover, they
generally disturb the state of the environment, and hence, the
correlations between $\cS$ and $\cE$. Unless all observers had agreed
to measure the environment in the same basis, their subsequent
measurements on the environment might not yield consensus about the
system, and one would not be able to attribute ``objective
properties'' to its state.

The solution to this puzzle becomes obvious after a close inspection
of how we learn about systems in the real world. Not only do the
independent observers gather information about $\cS$ indirectly by
measuring the environment, but different observers have access to
{\em disjoint} fragments of $\cE$.  By definition, when the same information
about $\cS$ can be discovered from different fragments, it must have
been imprinted in the environment {\em redundantly}. In addition, when
many disjoint fragments of the environment contain information about
the state of the system, its properties can be found out by different
observers without the danger of invalidating each other's
conclusions. This is because observables acting on disjoint fragments
of $\cE$ always commute with each other. Hence, the first two
requirements for objectivity are satisfied.

Moreover, and this is a crucial result of our study, we will
demonstrate that redundant imprinting in the environment selects a
preferred set of system observables. In
particular, for obvious reasons~\cite{WZ82a} the environment-promoted
amplification of information required to arrive at a redundant
imprinting cannot amount to cloning.  Amplification comes at the price
of singling out commuting system observables. As a consequence, even initially ignorant observers
performing arbitrary measurements on their fragments of $\cE$ will
find out only about this unique observable. Selectivity of
amplification establishes that redundant imprinting in the environment
is a sufficient requirement for the emergence of classical objective
reality.  Thus, the state of the system is {\it de facto} objective
when its complete and redundant imprint can be found in $\cE$.

Quantum Darwinism makes novel use of information theory by focusing on
the communication capacity of the environment. This approach
complements the conventional ``system-based'' treatments of
decoherence. There, the environment is above all an ``information
sink'', a source of decoherence responsible for irreversible loss of
information~\cite{Zur81a, ZHP93a, GJK+96a, PZ01a}. However, these two
complementary approaches do agree in their conclusions: as we will
show, the pointer observables singled out by einselection are the only
ones that can leave a redundant imprint on the
environment. In part, this can be understood as a
consequence of the ability of the pointer states to persist while
immersed in the environment. Moreover, in both approaches {\em
predictability} of the preferred states of the system (i.e., either
from initial conditions, or from many independent observations on
$\cE$) is the key criterion.  This predictability is tied to the
resilience that allows the information about the pointer observables
to proliferate, very much in the spirit of the ``survival of the
fittest'', and corroborates conjectures about the role of quantum
Darwinism in the emergence of objectivity we have described before
\cite{OPZ04b, Zur03a, Zur03b}.

\section{Definitions and conventions}
\label{sec:defs}

\begin{figure}[tp]
\includegraphics[width=2.7in]{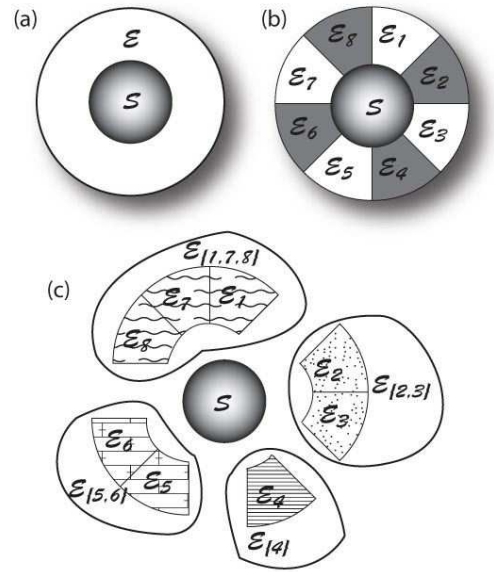}
\caption{The role and the structure of the environment in decoherence and in
quantum Darwinism.  (a) Decoherence treats environment as a monolithic entity
inaccessible to observers. The only role of $\cE$ is to be a ``sink'' of information 
about $\cS$. This is enough to understand einselection --- the emergence of 
the preferred pointer observable in the system, with decoherence - free pointer 
eigenspaces. In our Universe (b) environments are typically {\it not} monolithic. 
Rather, they consist of subsystems, e.g. atoms or photons. From the point of view 
of decoherence, there is no difference between the paradigms illustrated by (a) 
and (b): The same environment will lead to the same evolution of the system (e.g., in 
quantum Brownian motion the same environment can be modeled either as a monolithic 
field or as a collection of harmonic oscillators). Quantum Darwinism (c) recognizes 
that observers acquire their information about $\cS$ without interacting with it
directly -- from the imprints left by $\cS$ on the fragments of $\cE$. Such fragments are 
generally much smaller than $\cE$. The central result of this paper is to show 
that the selected information about the system that is inscribed redundantly
--- in many copies --- on the environment, and can be therefore found out independently 
from different fragments of $\cE$ by many observers, concerns the 
einselected pointer observable.}
\label{fig:partition}
\end{figure}

In the setting we are considering, a system $\cS$ with Hilbert space
$\cH^\cS$ interacts with an environment $\cE$ with Hilbert space
$\cH^\cE$. We denote the dimension of these state spaces by $d^\cS$
and $d^\cE$ respectively.  Furthermore, we assume that the environment
is composed of $N$ environmental subsystems $\cE_1, \cE_2,\ldots
\cE_N$ (see Fig.~\ref{fig:partition}). That is, its Hilbert space has
a {\em natural} tensor product structure $\cH^\cE = \bigotimes_{k=1}^N
\cH^{\cE_k}$. This partition plays an important role in our analysis,
as it suggests a natural definition of the independently accessible
fragments of $\cE$. We will comment on it at the end of
Section~\ref{sec:info}.

The joint quantum state of $\cS$ and $\cE$ is described by the density
operator $\rho^{\cS\cE}$ defined on $\cH^\cS \otimes \cH^\cE$. The
{\em reduced state} of the system is obtained by ``tracing out'' the
environment $\rho^\cS = \Tr_\cE \{\rho^{\cS\cE}\}$. It will often be
useful to consider the joint state of the system and {\em a fragment}
of the environment. Such fragment --- i.e. a subset of $\cE =
\{\cE_1,\cE_2,\ldots,\cE_N\}$ --- will be denoted by $\cF$. The
reduced state of $\cS$ and $\cF$ is obtained by tracing out the
complement of $\cF$: $\rho^{\cS\cF} = \Tr_{\overline \cF}
\{\rho^{\cS\cE}\}$, where $\overline\cF = \cE - \cF$.

Following textbook quantum mechanics, we call ``observable of $\cS$''
(resp. of $\cE$) any Hermitian operator defined on $\cH^\cS$
(resp. $\cH^\cE$). By convention, we will use the first letters of the
alphabet $\bA,\bB,\bC$ to denote {\em system} observables while the
last letters $\bX,\bY,\bZ$ will be reserved for {\em environment}
observables. Hermitian operators can be written in their spectral
decomposition, e.g.:
\begin{equation}
\bA = \sum_j a_j A_j.
\end{equation}
Adding to our convention, observables are denoted by bold capital
letters, their eigenvalues by lowercase letters, and spectral
projectors by capital letters. Only the spectral projectors are of
interest to us as they completely characterize the measurement
process, and the correlations between observables. We note that as we
shall deal with Hermitean observables, coherent states that are the
approximate pointer states in many situations of interest (e.g, in
underdamped harmonic oscillators) are beyond the scope of our study.

We will use the words ``system'' and ``environment'' in a very broad
sense. Without loss of generality, we will suppose that $\cH^\cS$ is
the part of the Hilbert space of $\cS$ containing the degrees of
freedom of interest. Even when the system is macroscopic, e.g. a
baseball, we are typically only interested in a few of its degrees of
freedom, e.g. center of mass, local densities, etc. Moreover, the
degrees of freedom of $\cS$ that do not couple to $\cE$ (directly or
indirectly) play no role in our analysis. Hence $d^\cS$ can remain
reasonably small even for fairly large systems: $d^\cS$ is really the
number of relevant distinct physical configurations of $\cS$. This
considerably simplifies the notation without compromising the rigor of
our analysis.

Similarly, it is not necessary to incorporate ``all the rest of the
Universe'' in $\cE$. $N$ really counts the number of environmental
subsystems that may have been influenced by the system: only they may
contain information about $\cS$. Hence, in many situations --- such as
a photon environment scattering off an object --- the ``size'' of
relevant $\cE$ can grow over time.

\section{Information}
\label{sec:info}

The approach to classicality outlined above is based on the existence
of correlations between $\cS$ and its environment that can be
exploited by various observers to find out about the system. As both
$\cS$ and $\cE$ are quantum systems, quantum information theory may
appear to be the right tool to study these correlations. This avenue
has been considered in the past \cite{Zur00a,Zur03a} in parallel with
the approach we pursue here and in \cite{OPZ04b} and is currently also
under investigation~\cite{BKZ04a}.  However, the emphasis here is on
the observables, and information about them is easier to characterize
through the relevant hypothetical measurements.  As in \cite{OPZ04b},
we focus on what can actually be found out about various observables
of the system by monitoring observables in fragments of the
environment.

The core question we ask is: {\em how much does one learn about
observable $\bA$ by measuring a different observable $\bX$?} This
question has an operational meaning. An observer may be considering
measurement of the observable $\bA$, but cannot predict its outcome
with certainty. To reduce his ignorance, he can choose to measure a
{\em different} observable $\bX$. By doing so, he may decrease his
uncertainty about the value of $\bA$. The amount by which his
uncertainty decreases is precisely the information gain we are going
to study. It represents the average number of {\em bits} required to
write down, in the most efficient way, the relevant data about $\bA$
acquired through the measurement of $\bX$.

We will mostly be interested in the case where $\bA$ acts on the
system and $\bX$ acts on a fragment of the environment --- which
automatically implies that $[\bA,\bX]=0$. However, we will
occasionally need to consider the correlations between the
measurements carried successively on the same system. Hence, we
present here the general case and return to the special case of
commuting observables in the next section. Thus, $\bA$ and $\bX$, with
spectral projectors $A_i$ and $X_j$, are arbitrary physical
observables acting on an arbitrary system, in the state described by
the density matrix $\rho$. In the following paragraph, there is no
environment, just one system and two observables that may or may
not commute.

The observer's uncertainty about the measurement outcome of $\bA$ is
given by the corresponding {\em Shannon entropy}:
\begin{equation}
H(\bA) = -\sum_i p(A_i)\ln p(A_i)
\end{equation}
where the probability associated to the measurement outcome ``$i$''
--- with the spectral projector $A_i$ --- is given by Born's rule
$p(A_i) = \Tr\{A_i\rho\}$. Entropy measures ignorance about the value
of $\bA$, the average number of bits missing to completely determine
its value. When the measurement of observable $\bX$ is performed and
outcome $X_j$ is obtained, the state of the system is updated to
\begin{equation}
\rho \xrightarrow{X_j} \rho _{|X_j} = \frac{X_j \rho X_j}{p(X_j)}
\label{eq:state_update}
\end{equation}
according to the projection postulate of quantum
theory~\cite{vN55a}. This state update changes the probability
assignment of the measurement outcomes of $\bA$:
\begin{equation}
p(A_i|X_j) = \Tr\{A_i \rho_{|X_j}\} = \frac{\Tr\{A_i X_j\rho
X_j\}}{p(X_j)}.
\label{eq:cond_prob}
\end{equation}
It is customary to call $p(A_i|X_j)$ ``the conditional probability of
$A_i$ {\em given} $X_j$'' and similarly, $\rho_{|X_j}$ is ``the
conditional state of the system {\em given} $X_j$''.

Thus, when $\bA$ is measured subsequently to $\bX$, the randomness of
its outcome would be characterized by:
\begin{equation}
H(\bA|X_j) = -\sum_i p(A_i|X_j)\ln p(A_i|X_j).
\end{equation}
The {\em conditional entropy} of $\bA$ given $\bX$ is the average of
this quantity over the measurement outcomes of $\bX$: $H(\bA|\bX) =
\sum_j p(X_j) H(\bA|X_j)$. The difference between the initial entropy
of $\bA$ and its entropy posterior to the measurement of $\bX$ defines
the {\em mutual information} we shall use throughout:
\begin{equation}
I(\bA:\bX) = H(\bA) - H(\bA|\bX).
\label{def:mutual}
\end{equation}
This is the average amount of information about $\bA$ obtained by
measuring $\bX$.

In quantum mechanics, it is possible that a certain measurement {\em
decreases} one's ability to predict the outcome of a subsequent
measurement, so mutual information is not necessarily positive. This
is in fact the reason why direct measurements on the system cannot be
used to arrive at a consensus about the state of the system. A direct
measurement by one observer will invalidate the knowledge acquired by
another when their measurements do not commute. However, this
disturbance can be avoided when the measurements are carried out on
different subsystems, since the observables commute
automatically. The mutual information between such observables has
extra properties that we shall now describe.

\subsection{Correlations between system and environment}
\label{sec:info_commute}

\begin{figure}[tp]
\centering  \includegraphics[width=5cm]{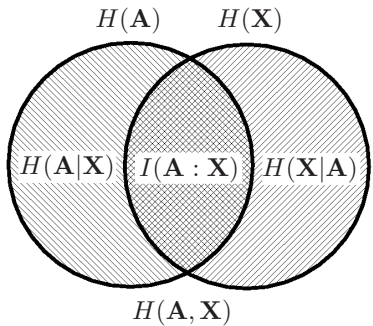}
\caption{Venn diagram for classical information. $H(\bA)$ is the
entropy associated with a measurement of $\bA$: it is the information
needed to completely determine its outcome. $H(\bX)$ is the same
quantity for the observable $\bX$.  $H(\bA,\bX)$ is the {\em joint}
entropy of $\bA$ and $\bX$. $H(\bA|\bX)$ is the average uncertainty
about $\bA$ remaining after a measurement of $\bX$. The information
learned about $\bA$ by measuring $\bX$ is thus $I(\bA:\bX) = H(\bA) -
H(\bA|\bX)$. The equivalent definition $I(\bA:\bX) = H(\bA) + H(\bX) -
H(\bA,\bX)$ can also be understood from the diagram. These two
definitions of the mutual information are equivalent \cite{CT91a} when
probabilities can be consistently  assigned to the outcomes of all the
possible measurements of $\bA$ and $\bX$  both separately and
jointly: this is ensured when $\bA$ and $\bX$ commute. They need not coincide otherwise.}

\label{Info}
\end{figure}

Let us now consider the case where $\bA$ acts on $\cS$ and $\bX$ on
$\cE$, or on a fragment $\cF$ of $\cE$. As $[\bA,\bX]=0$, the order in
which the measurements are carried out does not change the joint
probability distribution $p(A_i,X_j) = \Tr\{A_i X_j\rho X_j\} =
\Tr\{X_j A_i \rho A_i\}$. It follows from
Eqs.~(\ref{eq:state_update}-\ref{def:mutual}) that the mutual
information defined above can be written in an explicitly symmetric
form:
\begin{eqnarray}
I(\bA:\bX) &=& \sum_{ij} p(A_i,X_j)
\ln\frac{p(A_i,X_j)}{p(A_i)p(X_j)}\\ 
&=& H(\bA) + H(\bX) - H(\bA,\bX)
\label{eq:mutual_sym}.
\end{eqnarray}
The amount of information about $\bA$ that is obtained by measuring
$\bX$ is equal to the amount of information gained about $\bX$ by
measuring $\bA$, and is always positive. In this special
case, there is a nice diagrammatic representation of the information theoretic
quantities, shown on Fig.~\ref{Info}. 

Of particular interest to us is the {\em maximum} amount of
information about observable $\bA$ of $\cS$ that can be gained by
measuring a fragment $\cF$ of the environment. This is defined as \begin{equation}
\hat I_{\cF}(\bA) = \max_{\bX \in \mathcal M_\cF} I(\bA:\bX)
\end{equation}
where $\mathcal M_\cF$ is the set of measurements acting on $\cF$
only, i.e. the set of Hermitian operators that act trivially on
$\cH^{\overline\cF} = \bigotimes_{\cE_k \in \overline\cF} \cH^{\cE_k}$. 
The ``hat''  above is to emphasize that it is the maximum amount of information.

In particular, the maximum amount of information about the system
observable $\bA$ that can be retrieved from the {\em entire}
environment is denoted by $\hat I_\cE(\bA)$. This quantity plays a
crucial role in our analysis as only when $\hat I_\cE(\bA) \approx
H(\bA)$ can we hope to ``find out'' about $\bA$ by probing the
environment: the amount of information in the environment, $\hat
I_\cE(\bA)$, must be sufficiently large to compensate for the
observer's initial ignorance, $H(\bA)$, about the value of $\bA$.

\subsection{Redundancy of information in the environment}
\label{sec:redundancy}

When $\hat I_\cE(\bA) \approx H(\bA)$, the value of $\bA$ can be found
out indirectly by probing the environment. However, as noted in
Section~\ref{sec:motivation}, for many observers to arrive at a
consensus about the value of an observable $\bA$, there must be many
copies of this information in $\cE$. As a consequence, independent
observers will be able to perform measurements on disjoint subsets of
the environment, without the risk of invalidating each other's
observations.

{\em Redundancy} is therefore defined as the number of disjoint
subsets of the environment containing almost all --- all but a
fraction $\delta$ --- of the information about $\bA$ present in the
entire environment. Formally, let $\cF_1,\cF_2,\ldots,\cF_R$ be $R$
disjoint fragments of the environment, $\cF_i \cap \cF_j = \emptyset$
for $i\neq j$. Then,
\begin{eqnarray}
&& R_\delta(\bA) \label{eq:redundancy}\\ 
&& = \max_{\{\cF_j\}} \left\{R: \hat I_{\cF_j}(\bA) \geq (1-\delta)\hat I_\cE(\bA),\ \forall
j=1,\ldots,R \right\} \nonumber
\end{eqnarray}
where the maximization is carried over all partitions of
$\{\cE_1,\cE_2,\ldots \cE_N\}$ into disjoint subsets. Clearly, for any
observable $\bA$, $1\leq R_\delta(\bA) \leq N$. Redundancy
simply counts the number of copies of the imprint of $\bA$ in $\cE$, and
hence the maximum number of observers that can independently 
find out about $\bA$ from $\cE$.

\subsection{Fragments of the environment and elementary subsystems}

Before closing this section, we wish to emphasize the distinction we
are making between fragments of $\cE$ and elementary subsystems, and
comment on the role they play in using environment as a witness. The elementary
subsystems of $\cE$ are defined through the natural tensor product
structure of $\cH^\cE$. We assume this structure to be given and
fixed. In the case of a photon environment for instance, an elementary
subsystem could consist of a single photon. A fragment of $\cE$ on the
other hand is a collection of such elementary subsystems. For example,
while no single photon can reveal the position of an object, a small
collection of them, say 1000, may be enough to do so.

The optimization over {\em partitions} of the environment appearing at
Eq.~(\ref{eq:redundancy}) is necessary to arrive at a proper
mathematical definition of redundancy as there is no {\em a priori}
preferred partition. This will allow us to derive very general
consequences of redundancy, at the price of some technical (and
perhaps also conceptual) complications. However, for the purposes of
the emergence of a consensus among several observers --- in essence an
operational objective reality --- this partition should reflect the
distinct fragments of environment accessible to the different
observers. While our results hold for any such partition of the
environment into disjoint subsystems, Nature ultimately 
determines what part of $\cE$ is available to each observer.

The entire environment as a witness approach --- and more precisely
the very concept of redundancy --- capitalizes on the fact that the
environment has a tensor product structure $\cH^\cE =
\bigotimes_{\cE_k}\cH^{\cE_k}$. This raises the obvious question ``who
decides what are the elementary subsystems of $\cE$?''  Our primary
concern here is to provide a mechanism by which quantum systems can
exhibit ``objective existence'', the key symptom of the classical
behavior. As we will demonstrate, this can be achieved given that the
environment has a partition into subsystems, {\em regardless of what
these subsystems are}. Hence, what really matters is that {\em various
observers monitor different fragments of the environment}. As long as
this requirement (which guarantees that the observables they measure
commute) is fulfilled, their definition of subsystems of $\cE$ need
not coincide.  Section~\ref{sec:theory} will present unavoidable
consequences of this fact, without paying attention to the definition
of the environmental subsystems. Thus, our most important conclusion,
--- that redundancy implies selection of preferred observables --- is
independent of any particular choice of a tensor product in the
environment.

However, different tensor product structure of the environment can
{\em a priori} yield different redundantly imprinted observables since
redundancy itself makes reference to the tensor product structure.
There is no definite answer to what defines an elementary
environmental subsystem, but some considerations point towards
``locality'' as a judicious guideline.  For instance, particles are
conventionally defined by the symmetries of the fundamental
Hamiltonians of Nature, that are {\em local}. When we choose the
particles of the standard model as the elementary subsystems, we are
naturally led to local couplings between $\cS$ and $\cE$.  They will
determine how the information is inscribed in $\cE$. After all,
``there is no information without representation''. Moreover, the
information acquisition capacities of the observers are also
ultimately limited by the fundamental Hamiltonians of Nature. The
different observers occupy, and therefore monitor, different spatial
regions. Therefore, the monitored fragments $\cF_i$ entering in the
definition of redundancy --- as well as the elementary subsystems
$\cE_k$ composing them --- should reflect these distinct spatial
regions.

The fact that some division of the Universe into subsystems is needed
has been pointed out before. Indeed, the measurement problem
disappears when the Universe cannot be divided into subsystems
\cite{Zur93b, Zur03a,Sch03a}. Therefore, assuming that such division
exists in the discussion of the information-theoretic aspects of the
origin of the classical does not seem to be a very costly assumption.

\section{Consequences of redundancy}
\label{sec:theory}

We now have all the necessary ingredients to study the consequences of
the existence of redundant information about the system in the
environment. Here, we derive several properties of the system's
redundantly imprinted observables, as well as properties of the
environmental observables revealing this information. While each of
these results is interesting in its own right, our ultimate goal is to
combine them and to show that the redundancy singles out a preferred
set of commuting observables, the already familiar pointer
observables. Throughout this section, we
assume the existence of a {\em perfect} and {\em redundant} record of
the system observables $\bA,\bB,\bC,\ldots$ in the environment,
i.e. $\hat I_\cE(\bA) = H(\bA)$ and $R_{\delta = 0}(\bA) \gg 1$, and
similarly for $\bB,\bC,\ldots$ The general case of imperfect imprints
will be addressed in the next section. Let us begin by studying the
consequences of the existence of a record about the value of $\bA$ in
the environment.

\begin{lemma} \label{lemma:duality}
$\hat I_\cF(\bA) = H(\bA)$, iff there exists an observable $\bX' \in
\mathcal M_\cF$ for which $H(\bA|\bX') = 0$ and $H(\bX'|\bA) = 0$. The
measurement of $\bX'$ on a fragment of the environment reveals all the
information of the system observable $\bA$, and vice versa.
\end{lemma}

\begin{proof}
When observable $\bA$ is completely encoded in a fragment of the
environment, $\hat I_\cF(\bA) = H(\bA)$, there exists $\bX \in
\mathcal M_\cF$ for which $I(\bA:\bX) = H(\bA)$, which implies $H(\bA|\bX)=0$. As noted in
Section~\ref{sec:info_commute}, the mutual information between $\bA$
and $\bX$ is symmetric when these observables act on distinct systems,
i.e. $\cS$ and $\cF$. Therefore, measuring $\bA$ directly on the
system provides an amount of information $I(\bA:\bX) = H(\bA)$ about
the value of $\bX$, thereby decreasing its entropy to $H(\bX | \bA) =
H(\bX) - H(\bA)$. In general, this is not {\em all} the information
about $\bX$, as $H(\bX)$ may be larger than $H(\bA)$: $\bX$ reveals
all of the information about $\bA$ but the reverse is in general not
true. However, by picking a suitable coarse graining $\bX'$ of $\bX$,
it is always possible to establish the duality $H(\bA|\bX') = 0$ and
$H(\bX' | \bA) = 0$.

This can be seen quite simply. The equality $H(\bA|\bX) = 0$ implies
that given $X_j$, $A_i$ is determined: each measurement outcome of
$\bX$ points to a unique measurement outcome of $\bA$. This defines a
map $f: X_j \rightarrow A_i$. Such map may not be one-to-one, so the
conditional probability $p(X_j|A_i)$ of $X_j$ given $A_i$ is not
necessarily deterministic. However, we can construct the {\em coarse
grained projectors}
\begin{equation}
X_i' = \sum_{j:f(X_j) = A_i} X_j
\end{equation}
by regrouping the $X_j$ in the pre-image of $A_i$. The outcome of the
associated measurements $\bX'$ are therefore in one-to-one
correspondence to the outcomes of $\bA$, yielding the stated
duality. The converse is trivial with $X = X'$.
\end{proof}

An important corollary can be derived from Lemma~\ref{lemma:duality}
and the following observation: when the outcome of a projective
measurement on a system is deterministic, the act of measuring does
not modify the state of the system.

\begin{corollary}\label{corollary:effect}
Measurements of $\bA$ and $\bX'$ have exactly the same effect on the
joint state of the system and the environment:
\begin{equation}
X_j' \rho^{\cS\cE} X_j' = A_jX_j'\rho^{\cS\cE}X_j'A_j = A_j
\rho^{\cS\cE} A_j,
\label{eq:commute}
\end{equation}
which implies $\rho^{\cS\cE}_{|\tilde X_j} = \rho^{\cS\cE}_{|A_j}$. 
\end{corollary}

The contents of Lemma~\ref{lemma:duality} and
Corollary~\ref{corollary:effect} formalize our intuitive understanding
of the existence of a perfect record of the information about $\bA$ in
$\cE$: it allows perfect {\em emulation} of the direct measurement
$\bA$ by the indirect measurement $\bX'$. This emulation is not only
perfect from the point of view of its information yield, it also has
the same physical effect on the state of $\cS\cE$. This can be
regarded as an illustration of the fact that ``information is
physical'' \cite{Lan91a}: whenever the same information can be
retrieved by two different means, the disturbance on the quantum state
{\em can} be identical.\footnote{In the theory of generalized
measurements \cite{Kra83a}, one can in principle make a measurement
that yields no information whatsoever, yet greatly disturb the system;
hence the emphasized ``can'' in the above sentence.} Here, the
measurement of $\bX$ can in principle yield {\em more} information
than the direct measurement $\bA$ itself, e.g. $\bA$ may be a very
coarse grained observable. This is why it is in general necessary to
coarse grain $\bX$ in order to obtain equivalent information, and
therefore the same physical effect.

\begin{lemma}\label{lemma:rho_commute}
When $\hat I_\cF(\bA) = H(\bA)$, the observable $\bA$ commutes with
the reduced density matrix of the system, $[\rho^\cS ,\bA]=0$.
\end{lemma}

\begin{proof}
Following Lemma~\ref{lemma:duality} and
Corollary~\ref{corollary:effect}, there exists $\bX' \in \mathcal \cE$
for which $X_j' \rho^{\cS\cE} X_j' = A_j \rho^{\cS\cE} A_j$. The
reduced state of the system can be therefore written as
\begin{eqnarray}
\rho^\cS &=& \Tr_\cE\left\{\rho^{\cS\cE}\right\} \\ 
&=& \sum_j\Tr_\cE\left\{X_j'\rho^{\cS\cE}X_j'\right\} \\ 
&=& \sum_j\Tr_\cE\left\{A_j\rho^{\cS\cE}A_j\right\} =
\sum_jA_j\rho^{\cS}A_j,
\end{eqnarray}
so $\rho^\cS$ is block diagonal in the eigensubspaces of $\bA$.
\end{proof}

The relation between decoherence and the existence of an environmental
record has been pointed out and investigated in the past \cite{Zur81a,
Zur82a, Zur83a, Zur93b,Zur00a,GH97b,Zur98a,Hal99b}.  The above lemma
confirms that whenever the environment acquires a single copy of the
information about $\bA$, the state of $\cS$ decoheres with respect to
the spectral projectors $A_i$ of $\bA$. This decoherence-induced
diagonalization of the reduced state has additional important
implications when the information about the system is encoded {\em
redundantly} in the environment.

\begin{corollary}\label{corollary:induced}
Let $\bA$ be a system observable {\em redundantly} imprinted in $\cE$,
and let $\bX'$ be the coarse grained observable of the fragment $\cF$
of the environment that contains all the information about $\bA$ (as
constructed in Lemma \ref{lemma:duality}). Then the reduced state
$\rho^{\cS\cF}$ commutes with both $\bA$ and $\bX'$: the correlations
between these observables are classical.
\end{corollary}

\begin{proof}
By assumption $\bA$ is redundantly imprinted in $\cE$. Thus, its value
can be inferred by independent measurements, say $\bX$ and $\bY$,
acting on two distinct fragments of the environment, $\cF$ and
$\overline\cF = \cE - \cF$ respectively. Following Lemma
\ref{lemma:duality}, we can construct a coarse grained observable
$\bX'$ which retains essential correlations to $\bA$ and with the
property that $H(\bX' | \bA) =0$ --- a measurement of $\bA$ on $\cS$
reveals all the information about $\bX'$. As a consequence, the
measurement $\bY$ on $\overline\cF$ that reveals all the information
about $\bA$ must also reveals all the information about $\bX'$. Hence,
Corollary \ref{corollary:induced} follows from Lemma
\ref{lemma:rho_commute}.
\end{proof} 

We will now characterize the class of system observables that can be
{\em redundantly} imprinted in the environment. When $\bA$ and $\bB$
are redundantly imprinted in $\cE$ and redundancy is sufficiently
high, their value can be inferred from disjoint fragments of the
environment. This implies the following relation between any
redundantly imprinted observables.

\begin{lemma}\label{lemma:commute}
Let $\cF$ be a fragment of the environment containing a copy of the
information about $\bA$. When the rest of the environment
$\overline\cF$  contains a copy of the information about $\bB$, then
$\bA$ and $\bB$ {\em necessarily commute on the support of}
$\rho^\cS$.
\end{lemma}

\begin{proof}
The assumptions of Lemma~\ref{lemma:commute} imply the existence of two
environmental observables $\bX$ and $\bY$ acting on disjoint fragments
of the environment $\cF$ and $\overline\cF$ for which $H(\bA|\bX)=0$
and $H(\bB|\bY)=0$. Following the procedure of
Lemma~\ref{lemma:duality}, we can coarse grain these two observables
into $\bX'$ and $\bY'$ that are in one-to-one correspondence with
$\bA$ and $\bB$ respectively. The observables $\bX'$ and $\bY'$
obviously commute with each other as they act on disjoint fragments of
$\cE$, and also commute with the system observables $\bA$ and
$\bB$. Then, by Corollary~\ref{corollary:effect}, we have
\begin{eqnarray*}
A_iB_j\rho^{\cS\cE}B_jA_i &=& A_iY_j'\rho^{\cS\cE}Y_j'A_i =
Y_j'A_i\rho^{\cS\cE}A_iY_j'  \\ &=& Y_j'X_i'\rho^{\cS\cE}X_i'Y_j' =
X_i'Y_j'\rho^{\cS\cE}Y_j'X_i' \\ &=& X_i'B_j\rho^{\cS\cE}B_jX_i' =
B_jX_i'\rho^{\cS\cE}X_i'B_j \\  &=& B_jA_i \rho^{\cS\cE}A_iB_j,
\end{eqnarray*}
proving the lemma.
\end{proof}

The above Lemma states that it is impossible to acquire perfect
information about two non-commuting observables by measuring two
disjoint fragments of $\cE$ simultaneously. This is reminiscent of Heisenberg indeterminacy
principle. However, in spite of this similarity, these two results
differ in the precise setting in which they hold. Heisenberg principle
asserts that it is impossible to know simultaneously the values of two
non-commuting observables of an otherwise isolated system. In our
Lemma, the system is already correlated with its environment, and the
observer is trying to find out about the value of two observables {\em
using these correlations}. In other words, Heisenberg indeterminacy
concerns information about the system before it interacted with $\cE$,
while Lemma~\ref{lemma:commute} focuses on the information about $\cS$
that is present in $\cE$ after their interaction.

This difference illustrates the nature of our operational approach: we
are describing the physical properties of an open quantum system,
hence we focus on information about its present state, not on what it
was prior to the interaction with $\cE$.  Decoherence happens, so we
are dealing with it.\footnote{To fully grasp the distinction between
these two settings requires a somewhat technical discussion. Assume
that the system and environment are initially in the uncorrelated
state $\rho^{\cS\cE}(0) = \rho^\cS(0)\otimes\rho^\cE(0)$. They
interact for a time $t$, yielding the joint state $\rho^{\cS\cE}(t) =
U^{\cS\cE}\rho^{\cS\cE}(0) U^{\cS\cE\dagger}$ where $U^{\cS\cE} =
\exp\{-i{\bf H}^{\cS\cE}t\}$. When a measurement $\bX$ is carried on
the environment and outcome $j$ is observed, the conditional state of
the system will be
$$\rho^\cS_{|X_j}(t) = \frac{\Tr_\cE\left\{X_j \rho^{\cS\cE}(t)
X_j\right\}}{p(X_j)}.$$ According to the generalized theory of
measurement \cite{Kra83a}, this can be described as a positive
operator valued measure (POVM) acting on the {\em initial} state of
the system, $\rho^\cS(0)$. Formally, there exists a set of operators
$A_{ij}$ acting on $\cH^\cS$ such that $\rho^\cS_{|X_j}(t) = \sum_i
A_{ij} \rho^\cS(0) A_{ij}^\dagger$.  However, it does not correspond
to any kind of measurement acting on $\rho^\cS(t)$, the state of the
system {\em after} it has interacted with $\cE$ (see for example
\cite{Fuc02c} Eq.~(100)). Our Lemma applies also to this type of
information gathering processes that cannot be described within the
POVM formalism.}

We can derive a similar result for the commutation of environmental
observables.
\begin{lemma}\label{lemma:XYcommute}
Let $\cF$ be a fragment of the environment containing information
about both $\bA$ and $\bB$, two {\em commuting} system
observables. Then, there exist two observables $\bX'$ and $\bY' \in
\mathcal M_\cF$ that commute on the support of $\rho^{\cS\cE}$ and
reveal all the information about $\bA$ and $\bB$ respectively.
\end{lemma}

\begin{proof}
The assumptions of Lemma~\ref{lemma:XYcommute} imply the existence of
$\bX$ and $\bY \in \mathcal M_\cF$ such that $I(\bA:\bX) = H(\bA)$ and
$I(\bB:\bY) = H(\bB)$. Following the procedure of
Lemma~\ref{lemma:duality}, $\bX$ and $\bY$ can be coarse grained to
$\bX'$ and $\bY'$ while retaining all the essential correlations to
$\bA$ and $\bB$. Since $\bA$ and $\bB$ commute, the proof of
Lemma~\ref{lemma:commute} can be applied here, yielding the desired
result.
\end{proof}

Given these properties of redundantly imprinted observables, we can
now state a very important result, which essentially ensures the
uniqueness of redundantly imprinted observables.
\begin{theorem}\label{th:uniqueness}
Let $\bB$ and $\bC$ be two system observables redundantly imprinted in
the environment, and let $\cF_1,\cF_2,\ldots,\cF_R$ be $R\geq 2$
disjoint fragments of $\cE$, each containing a copy of the information
about $\bB$ and $\bC$. Then, there exists a {\em
refined} system observable $\bA$ satisfying:
\begin{enumerate}
\item $\hat I_\cE(\bA) = H(\bA)$,
\item $R_0(\bA) \geq R$,
\item $H(\bB|\bA) =0$ and $H(\bC|\bA) = 0$.
\end{enumerate}
\end{theorem}

\begin{proof}
The assumptions of the theorem imply the existence of $\bY^{\cF_n}$ and
$\bZ^{\cF_n} \in \mathcal M_{\cF_n}$ for which $H(\bB|\bY^{\cF_n}) =
0$ and $H(\bC|\bZ^{\cF_n}) = 0$ for all $n=1,\ldots,R$. By
Lemma~\ref{lemma:commute}, $\bB$ and $\bC$ must commute as they can be
inferred from disjoint fragments of the environment. This fact,
together with Lemma~\ref{lemma:XYcommute}, implies that there exists
{\em commuting} coarse grained observables $\bY^{\prime\cF_n}$ and
$\bZ^{\prime\cF_n} \in \mathcal M_{\cF_n}$ that reveal all of the
information about $\bB$ and $\bC$ respectively. The observables $\bB$
and $\bC$ can be merged into a more {\em refined} observable $\bA$
with spectral projectors given by the product of the spectral
projectors of $\bB$ and $\bC$:
\begin{equation}
A_{ij} = B_iC_j.
\end{equation}
As $B_i$ and $C_j$ commute, the operators $A_{ij}$ form a complete set
of mutually orthogonal projectors. Similarly, the environment
observables $\bY^{\prime\cF_n}$ and $\bZ^{\prime\cF_n}$ can be merged
into a more refined observable $\bX^{\cF_n}$ with spectral projectors
$X^{\cF_n}_{ij} = Y^{\prime\cF_n}_i Z^{\prime\cF_n}_j$. A measurement of
$\bX^{\cF_n}$ on $\cF_n$ reveals all the information about
$\bA$. Indeed, following Corollary~\ref{corollary:effect},
\begin{eqnarray*}
&& p(A_{i'j'}|X_{ij}^{\cF_n}) =
\frac{\Tr\left\{A_{i'j'}X_{ij}^{\cF_n}\rho^{\cS\cE}X_{ij}^{\cF_n}A_{i'j'}\right\}}
{\Tr\left\{X_{ij}^{\cF_n}\rho^{\cS\cE}X_{ij}^{\cF_n}\right\}}
\\ && \quad =
\frac{\Tr\left\{B_{i'}C_{j'}Y^{\prime\cF_n}_iZ^{\prime\cF_n}_j\rho^{\cS\cE}Z^{\prime\cF_n}_jY^{\prime\cF_n}_iC_{j'}B_{i'}\right\}}
{\Tr\left\{Y^{\prime\cF_n}_iZ^{\prime\cF_n}_j\rho^{\cS\cE}Z^{\prime\cF_n}_jY^{\prime\cF_n}_i\right\}}\\
&& \quad = \delta_{ii'}\delta_{jj'}
\end{eqnarray*}
which implies $H(\bA|\bX^{\cF_n}) =0$. This automatically implies
$\hat I_{\cF_n}(\bA) = H(\bA)$ and, since the above holds for all
$n=1,\ldots R$, we have $R_0(\bA) \geq R$. Finally, note that $B_i =
\sum_j A_{ij}$ and $C_j = \sum_i A_{ij}$; $\bB$ and $\bC$ are obtained
by coarse graining $\bA$. Therefore, a measurement of $\bA$ reveals
all the information about $\bB$ and $\bC$, hence completing the proof.
\end{proof}

The
meaning of this Theorem is that whenever more than one observable can
be redundantly inferred from a fixed set of disjoint fragments of
$\cE$, they necessarily correspond to some coarse grained version of a
{\em maximally refined redundantly imprinted observable} $\bA$. The
{\em modus tollens} of this Theorem is also very enlightening. Given a
decomposition of $\cE$ into fragments and the associated maximally
refined redundantly imprinted observable $\bA$, the only observables
$\bB$ that can be completely and redundantly inferred from these
fragments of $\cE$ are those for which $I(\bB:\bA) = H(\bB)$. The
value of $\bB$ must be entirely determined by a measurement of the
maximally refined $\bA$. This is obviously a sufficient condition: if
a direct measurement of $\bA$ reveals all the information about $\bB$
and $\bA$ is redundantly imprinted in $\cE$, then so is
$\bB$. Theorem~\ref{th:uniqueness} shows that this requirement is also
necessary.

\begin{figure}[tp]
\includegraphics[width=2.9in]{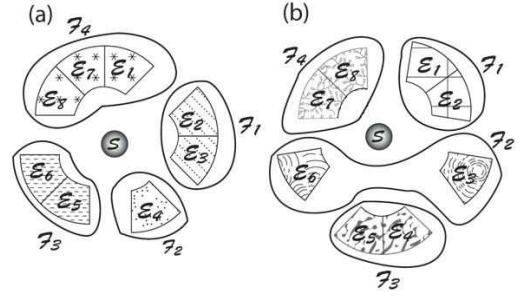}
\caption{Suppose $\bB$ and $\bC$ are redundantly imprinted in the
environment, but are associated respectively to decompositions (a) and
(b), represented by different arrangements of the subsystems of $\cE$
into fragments. Then, from
Theorem~\ref{th:uniqueness} one cannot yet conclude that there exists a
maximally refined observable $\bA$ from which both $\bB$ and $\bC$ can
be inferred. This is because the two decompositions do not
match. However, when the redundancy of $\bB$ and $\bC$ satisfies
$R_0(\bB)R_0(\bC) > N$, then Theorem~\ref{th:commute} proves that the
maximally refined observables obtained for each decomposition of $\cE$
into fragments must commute (i.e. they are compatible).}
\label{fig:2_partitions}
\end{figure}

Note that the proof relies only on a redundancy greater
than $2$ and on a fixed decomposition of $\cE$ into disjoint fragments
from which $\bB$ and $\bC$ are inferred. Relaxing this last
assumption raises the possibility that different decompositions
lead to incompatible maximally refined redundant observables.
Figure \ref{fig:2_partitions} represents two decompositions of $\cE$ into
fragments for which the corresponding maximally refined
redundantly imprinted observables could not be compared
through the above theorem. In absence of further assumptions
on the dynamics that is responsible for the creation of correlations
between $\cS$ and fragments of $\cE$, this possibility can be ruled out 
when redundancy is high (see Theorem \ref{th:commute} below). For most
practical cases---especially when the subsystems of $\cE$ interact
with $\cS$ according to commuting hamiltonians---such complication
can be avoided as the maximally refined observable is often
independent of the decomposition of $\cE$ into fragments.

\begin{theorem}
Let $\bB$ and $\bC$ be two system observables both highly redundantly
encoded in $\cE$, such that $R_0(\bB) \times R_0(\bC) > N$. Then,
$[\bB,\bC] = 0$ on the support of $\rho^{\cS}$.
\label{th:commute}
\end{theorem}

\begin{proof}
By the assumptions of the Theorem, there exists a fragment $\cF$ of
size at most $N/R_0(\bB)$ containing all information about
$\bB$. Since $R_0(\bC) > N/R_0(\bB)$, there exists a fragment $\cF'$
disjoint from $\cF$, i.e. $\cF \cap \cF' = \emptyset$, that contains
all information about $\bC$. By Lemma~\ref{lemma:commute} we have
$[\bB,\bC]=0$.
\end{proof}

It is intriguing to note that --- in view of the above discussion ---
large redundancy implies $R > \sqrt N$. Theorems~\ref{th:uniqueness}
and \ref{th:commute} prove our claim of
Section~\ref{sec:motivation}. Redundant spreading of information comes
at the price of singling out a preferred set of commuting observables:
those obtained by coarse graining maximally refined redundantly
observables.

So far, we have not considered any dynamical evolution of the system
and environment: we have focused on the correlations present at a
fixed time without paying attention to how they arise. We studied
consequences of regarding environment as a witness (and interrogating
it about the state of the system), but we haven't yet enquired about
the dynamics that allowed the environment to acquire this information
in the first place. Ultimately, it is the coupling Hamiltonian between
$\cS$ and $\cE$ which establishes these correlations. The coupling is
also responsible for the emergence of preferred system observables ---
the pointer observables --- that are least affected by the openness of
the system. Thus, as in the case of einselection of the pointer
states, the connection between the emergence of an objective reality
and the selection of preferred observables has to be ultimately traced
to the dynamics responsible for the ``environmental monitoring' of
$\cS$ by $\cE$ --- i.e., through dynamical considerations. When the
correlations between maximally refined observables and fragments of
the environment {\em persist in time}, Lemma~\ref{lemma:rho_commute}
and Theorems~\ref{th:uniqueness} \& \ref{th:commute} imply the
following conclusion.

\begin{corollary}\label{corollary:pointer} 
Highly redundantly imprinted observables are system's pointer observables.
\end{corollary}

In other words, under the idealized assumption of perfect records we
have demonstrated that only the already familiar pointer observables
can leave a redundant and robust imprint on
$\cE$. Corollary~\ref{corollary:pointer} can be understood as a
consequence of the ability of the pointer states to persist while
immersed in the environment. This resilience allows the information
about the pointer observables to proliferate, very much in the spirit
of the Darwinian ``survival of the fittest''. This, along with other
facts, will be illustrated on a dynamical model in the next section.

\section{Quantum Darwinism: Dynamical emergence of objectivity}
\label{sec:model}

It should by now be clear that the presence of redundant information
in the environment imposes severe physical constraints on the state of
the system. Surprisingly, the derivation of these important facts
required little physics --- most of them followed from
information-theoretic arguments, once again illustrating that
information is physical. One could sum up our conclusions by noting
that the environment can be a perfect witness --- with multiple
perfect records --- but only for one set of commuting observables
defined by the perfect maximally refined pointer
observables. Such
perfection will generally be only approximated in the real world. It
is therefore important to investigate the situation where multiple,
yet imperfect, copies arise through the dynamics. To this end we shall
consider selective proliferation of information in a simple model that
exhibits quantum Darwinism.

By focusing on the {\em dynamics} of $\cS\cE$ rather than on their
{\em state} we will be able to extend our analysis. In particular, we
will show that when the imprints of the objective observable are {\em
nearly perfect} --- i.e.\@~$\hat I_\cE(\bA) \geq (1-\vartheta) H(\bA)$
and $R_\delta(\bA) > 1$ for $\vartheta \ll 1$ and $\delta \ll 1$ {\em
finite} but small --- the conclusions obtained earlier still
hold. Above all, along the lines of Theorem~\ref{th:uniqueness}, there
exists a unique maximally refined observable whose information is the
only one available in fragments of $\cE$. Finally, we will be
able to specify the {\em optimal} measurement to be performed on
fragments of $\cE$ to learn about the system. This is a considerable
improvement over the result of the last section where, in the absence
of any model Hamiltonian, we could only demonstrate the existence of
such a measurement.

\subsection{Dynamics of Quantum Darwinism and decoherence}
\label{sec:model_dynamics}

The model we consider is a generalization of the simple early model of
decoherence put forward in \cite{Zur82a} and throughly investigated as
a tractable, yet non-trivial, paradigm for decoherence: despite its
simplicity, it captures the essence of einselection. Hence, our simple
model serves as a special case that sets a conceptual framework for
the study of more sophisticated models, such as a photon environment
scattering on an object and carrying away potential visual data. As
any specific model, it requires some fairly specific assumptions. We
shall consider their role in the next section and discuss the extent
to which relaxing some of these assumptions affects our conclusions.

A system $\cS$ is coupled to an environment $\cE = \{\cE_1, \ldots,
\cE_N \}$ through the Hamiltonian:
\begin{equation}
{\bf H}^{\cS\cE} = \sum_{\cE_k \in \cE} \bA \otimes \bZ^{\cE_k},
\label{eq:coupling}
\end{equation}
where $\bA$ and $\bZ^{\cE_k}$ are operators acting on $\cS$ and
$\cE_k$ respectively. The joint initial state of the system and the
different environmental subsystems is assumed to be a pure product
state:
\begin{equation}
\ket{\Phi^{\cS\cE}(0)} = \ket{\phi^\cS}\otimes
\ket{\phi^{\cE_1}}\otimes\ket{\phi^{\cE_2}}\otimes\ldots.
\end{equation}
Hence, before the interaction, there is no correlation between the
system and the environment, nor among the environmental
subsystems. 

A convenient way of writing $\ket {\phi^{\cS}}$ in view of expressing
the time evolution of the joint state of $\cS\cE$ is to decompose it
in an eigenbasis $\ket i$ of $\bA$ ($\bA \ket i = a_i \ket i$):
\begin{equation}
\ket{\phi^\cS} = \sum_i \alpha_i \ket i.
\label{eq:initial_state}
\end{equation}
Without loss of generality, we can assume that the vectors $\ket i$
with non-zero coefficients $\alpha_i$ in
Eq.\@~(\ref{eq:initial_state}) are associated with distinct
eigenvalues $a_i$ of $\bA$. This can be done by choosing appropriate
bases for the degenerate eigenspaces of $\bA$. Thus, after an
interaction time $t$ with the $N$ subsystems of the environment, the
joint state of $\cS\cE$ evolves into:
\begin{eqnarray}
\ket {\Phi^{\cS\cE}(t)} &=& \sum_j \alpha_j \ket j \bigotimes_{\cE_k
\in \cE} e^{-ita_j \bZ^{\cE_k} } \ket {\phi^{\cE_k}} \\ &=&  \sum_j
\alpha_j \ket j \bigotimes_{\cE_k \in \cE} \ket{\phi_j^{\cE_k}},
\label{eq:state_time}
\end{eqnarray}
where we have defined $\ket{\phi_j^{\cE_k}} = e^{-ita_j \bZ^{\cE_k} }
\ket {\phi^{\cE_k}}$.

In the following sections, we will analyze the correlations between
the system $\cS$ and the ``observed'' part of the environment
$\cF$. For example, in the case of a photon environment scattering on
an object, $\cF$ is the set of photons that hit the observer's retina,
while $\overline \cF$ represents those photons that have scattered on
$\cS$ but that are not intercepted by the
observer. Equation~(\ref{eq:state_time}) allows us to compute the
state of $\cS$ and $\cF$:
\begin{equation}
\rho^{\cS\cF} = \sum_{ij} \alpha_i\alpha_j^* \kb{i}{j} \otimes
\kb{\Phi_i^\cF}{\Phi_j^\cF} \times \gamma^{\overline \cF}_{ij},
\label{eq:state_real}
\end{equation}
where
\begin{equation}
\ket {\Phi^{\cF}_j} = \bigotimes_{\cE_k \in \cF}
e^{-ia_j\bZ^{\cE_k}t}\ket{\phi^{\cE_k}} = \bigotimes_{\cE_k \in \cF}
\ket{\phi_j^{\cE_k}} \ ,
\label{eq:def_phi}
\end{equation} 
and $\gamma^{\overline \cF}_{ij} = \prod_{\cE_k \in \overline \cF} \bk
{\phi^{\cE_k}_j}{\phi^{\cE_k}_i} =
\bk{\Phi_j^{\overline\cF}}{\Phi_i^{\overline\cF}}$. Similarly, the
reduced state of the system at time $t$ reads:
\begin{equation}
\rho^\cS = \sum_{ij} \alpha_i \alpha_j^* \kb i j \times
\gamma^\cE_{ij},
\label{eq:system_decoherence}
\end{equation}
where $\gamma^\cE_{ij} = \prod_{\cE_k \in \cE} \bra{\phi^{\cE_k}}
e^{-i t(a_i - a_j)\bZ^{\cE_k} }\ket{\phi^{\cE_k}}$. Above, the
$\gamma$'s are called the {\em decoherence factors}.

Except in carefully controlled experiments --- where the system can be
very well isolated from its environment --- the number of
environmental subsystems interacting with $\cS$ is huge. In this case,
the $\gamma_{ij}^\cE$'s will be typically very small for $i\neq
j$. Each environmental subsystem contributes a factor
$\bra{\phi^{\cE_k}} e^{-i (a_i - a_j)\bZ^{\cE_k} t}\ket{\phi^{\cE_k}}$
to $\gamma_{ij}^\cE$, so unless $\ket{\phi^{\cE_k}}$ is an eigenstate
of $\bZ^{\cE_k}$ --- in which case $\cE_k$ is effectively decoupled
from the system --- this strictly decreases the decoherence factor.
Hence, $\gamma_{ij}^\cE$ goes to $\delta_{ij}$ as $N$ increases. To be
more specific, when the initial states of the environmental subsystems
are distributed uniformly at random, the decrease of $\gamma^\cE_{ij}$
is exponential with $N$ and typically Gaussian with time \cite{ZCP03a}.
 
This reasoning also applies to the decoherence factors
$\gamma^{\overline \cF}_{ij}$: they tend to zero for $i \neq j$ as the
number of {\em unobserved} environmental subsystems increases. Even
though a considerable number of the environmental subsystems can be
intercepted by the observer, an even larger fraction will usually
escape his monitoring. Thus, the $\gamma_{ij}^{\overline \cF}$'s will
also typically be very small for $i\neq j$.

We have numerically studied a version of this model (the case of two
dimensional system and environmental subsystems) in \cite{OPZ04b}. Our
results indicate that for sufficiently large environment, several
system observables can become nearly perfectly imprinted in the
environment (see Fig. 1a of \cite{OPZ04b}). However, the results also
clearly show that only those observables very ``close'' to the pointer
observable can leave a {\em redundant} imprint in the environment (see
Fig. 1b of \cite{OPZ04b}). The information about the other system
observables can only be accessed by measuring the {\em entire}
environment: their value cannot be learned from small fragments of
$\cE$. In what follows, we will derive these facts analytically.

\subsection{Decoherence: the focus on the system}
\label{sec:conventional}

In order to contrast the environment as a witness approach to
classicality as well as to illustrate significance of the dynamics
that leads to quantum Darwinism, we will now review some results that
were obtained using more conventional approaches. Thus, we will
abandon for a moment the study of correlations between $\cS$ and parts
of the environment to focus uniquely on the state of the system, as it
is done in the the standard studies of decoherence \cite{GJK+96a,
PZ01a}.

We see from Eq.~(\ref{eq:system_decoherence}) that the off-diagonal
terms of the system's density matrix tend to be very small when
expressed in the $\ket i$ basis. It is therefore natural to expect
that these are the quasi-classical states of the system: the coupling
to the environment suppresses quantum superpositions of the states
$\ket i$. Of course, the {\em exact} instantaneous diagonal basis of $\rho^\cS$ might
differ from $\ket i$, especially when the
coefficients $\alpha_j$ of the system's initial state
Eq.~(\ref{eq:initial_state}) are almost equal. Hence, for someone
focused on the instantaneous state of the system, it may not be clear
why the basis $\ket i$ deserves any special attention in spite of the
existence of small --- yet non-zero --- off-diagonal elements for
$\rho^\cS$. However, a simple analysis shows that the basis $\ket i$
is the {\em only} basis for which all off-diagonal terms tend to zero
{\em independently} of the initial state of the system, and, thus,
retain correlations with the rest of the Universe. This ability to
retain correlations is behind the predictability of the pointer
states --- it is in fact their defining feature (see
\cite{Zur81a, Zur82a} for the initial formulation, and \cite{Zur03a,
Zur04a, Sch03a} for a recent re-assessment).  

Persistence of correlations implies continued existence of the states
\cite{Zur98a, Zur03a}. Independence of the pointer states from the
initial state of the system is obviously a very important property of
the to-be-classical states: the quasi-classical domain must be
independent of the initial state of the system --- e.g., the set of
the pointer states of the apparatus should not depend on the state of
the measured quantum system.  When the system is initially prepared in
one of the states $\ket i$, it will not be affected by the interaction
with the environment: pointer states are stable. In more general cases
(more complicated interactions, etc.), the above simple analysis
cannot be carried out, and one usually relies on the predictability
sieve \cite{Zur93b, ZHP93a, GJK+96a, PZ01a} to find the pointer
basis. Predictability sieve seeks most predictable initial states ---
states that produce the smallest amount of entropy over time while
subject to interaction with the environment.

\subsection{Perfect correlations}

We now return to the study of the correlations between $\cS$ and
fragments of the environment. Our goal is to characterize what kind
and how information is stored in the environment when it is
redundant. To be more precise, we will analyze the structure of
information in a fragment $\cF$ of $\cE$ under the assumption that the
rest of the environment, i.e. $\overline \cF = \cE - \cF$, contains at
least one additional copy of this information. Our demand that both $\cF$ and
$\overline \cF$ contain a copy of the information ensures a minimum
redundancy of $R \geq 2$ which --- although insufficient to ensure the
emergence of a consensus about the properties of $\cS$ among many
observers --- is enough to re-derive directly most of the results
established in Section~\ref{sec:theory}.
\footnote{In this toy model, all the environmental subsystems couple to the system in the same
way, so only the size of the fragments $\mathcal F$ and $\overline{\mathcal F}$ matters. Thus,
complications illustrated in Fig. \ref{fig:2_partitions} and analyzed in Theorem \ref{th:commute} do not occur, and redundancy $R \geq 2$ is indeed sufficient to establish uniqueness of the preferred observable.} In addition, we will be able
to specify optimal measurement strategies on $\cF$, the fragment of
$\cE$ accessible to the observer, for inferring any kind of
information about $\cS$.
 
To build our intuition, it is enlightening to first consider the
limiting case $\gamma^{\overline \cF}_{ij} = \gamma^\cF_{ij} =
\delta_{ij}$, where $\gamma_{ij}^\cF =
\bk{\Phi_i^\cF}{\Phi_j^\cF}$. We return to the general case in
Section~\ref{sec:model_imperfect}. Following the general argument of
Section~\ref{sec:model_dynamics}, perfect correlation will be
recovered when the environment is infinite. In our analysis, this
condition reflects the physical assumption that both the observed and
the unobserved fragments of $\cE$ are very large.  It is also possible
to obtain perfect correlations in much smaller systems when the
dynamics is fine-tuned so that the resulting state is GHZ-like. In
either case, the density matrix of $\cS$ and $\cF$ reads
\begin{equation}
\sigma^{\cS \cF} = \sum_i |\alpha_i|^2 \kb i i \otimes \kb
{\Psi^{\cF}_i} {\Psi^{\cF}_i}
\label{eq:classical_state}
\end{equation}
where the $\ket{\Psi^{\cF}_i}$'s are mutually orthogonal. Note that
$\rho$ and the $\ket{\Phi^\cF_i}$'s of
Eqs.~(\ref{eq:state_real}-\ref{eq:def_phi}) have been replaced
respectively by `$\sigma$ and $\ket {\Psi^\cF_i}$'s to emphasize
$\gamma^\cF_{ij} = \delta_{ij}$.

Consider the observable $\bX \in \mathcal M_\cF$ that perfectly
distinguishes between the orthogonal states $\ket{\Psi^\cF_i}$:
\begin{equation}
\bX = \sum_i x_i \kb{\Psi^\cF_i}{\Psi^\cF_i} = \sum_i x_i X_i .
\end{equation}
Following Eq.~(\ref{eq:state_update}), the state of the system after a
measurement of $\bX$ on $\cF$ with outcome $X_j$ is
\begin{equation}
\sigma^\cS_{|X_j} = \Tr_\cF\{\sigma^{\cS\cF}_{|X_j}\} =
\frac{\Tr_\cF\{X_j\sigma^{\cS\cF}X_j\}}{p(X_j)} = \kb jj.
\label{eq:example_update}
\end{equation}
This is the same as the state of $\cS$ after a {\em direct}
measurement of $\bA$ with outcome $A_j$, $\sigma^{\cS}_{|X_j} =
\sigma^{\cS}_{|A_j}$. Thus, a subsequent measurement of $\bA$ on the
system yields outcome $A_i$ with certainty, so $H(\bA|\bX) = 0$. The
same conclusion can be reached for a measurement acting on
$\overline\cF$: thus, $\bA$ is encoded redundantly in $\cE$ (there are
at least two copies). As a consequence, all the results of
Section~\ref{sec:theory} trivially hold (with exception of Theorem \ref{th:commute}). It is however instructive to
derive them directly for our specific model, without appealing  to the
general lemmas.

As seen from Eq.~(\ref{eq:example_update}), the {\em indirect}
measurement $\bX$ perfectly emulates the {\em direct} measurement
$\bA$ on the system. Consequently, all the information about the
observable $\bA$ can be extracted by the measurement $\bX$ on the
environment. In addition, we clearly see the ``no-information /
no-disturbance'' principle at work: once the outcome of $\bX$ is
known, measuring $\bA$ directly does not disturb the state of the
system any further, as the measurement of $\bX$ ``projects'' the
system in an eigenstate of $\bA$ (see
Corollary~\ref{corollary:effect}). Averaging over the measurement
outcomes of $\bX$ yields the reduced density matrix of $\cS$, which is
a mixture of the eigenstates of $\bA$, so $[\rho^\cS,\bA] =0$ as
specified by Lemma~\ref{lemma:rho_commute}.

Specifying the Hamiltonian responsible for the correlations between
$\cS$ and its environment allows one to go beyond the purely
information-theoretic analysis of Section~\ref{sec:theory}. Given the
explicit form of the density matrix $\sigma^{\cS\cF}$, we can address
more interesting questions such as: what is the optimal measurement
$\bY$ on $\cF$ that yields the largest amount of information about a
given system observable $\bB$? In fact, it turns out that $\bX$ is the
{\em optimal} measurement on $\cF$ to find out the value of {\em any}
system observable $\bB$. Regardless of what we wish to learn about the
system, the optimal strategy consists in measuring $\bX$ on $\cF$:
\begin{equation}
I(\bB:\bX) \geq I(\bB :\bY)
\label{eq:measure_opt}
\end{equation} 
for all $\bB \in \mathcal M_\cS$ and $\bY \in \mathcal
M_\cF$. Similarly, when a non-optimal measurement $\bY$ is carried on
$\cF$, it is always primarily correlated with the pointer observable
$\bA$. In other words, when a measurement $\bY$ is carried on a
fragment of the environment, it always reveals more information about
the pointer observable $\bA$ than any other system observable:
\begin{equation}
I(\bA:\bY) \geq I(\bB :\bY)
\label{eq:observable_opt}
\end{equation} 
for all $\bB \in \mathcal M_\cS$ and $\bY \in \mathcal M_\cF$.

Let us prove these two very important assertions. First, note that
following Eqs.~(\ref{def:mutual},\ref{eq:mutual_sym}), the mutual
information between any system observable $\bB$ and environmental
observable $\bY$ takes the following form:
\begin{eqnarray}
I(\bB:\bY) & = & H(\bB) - H(\bB|\bY) \\ &=& H(\bY) - H(\bY|\bB).
\end{eqnarray}
When $\bB$ is a {\em fixed} system observable, maximizing $I(\bB:\bY)$
amounts to minimizing $H(\bB|\bY)$. To do this, consider the
conditional state $\sigma^{\cS}_{|Y_i}$ of the system given a
measurement outcome $Y_i$ on $\cF$. Simple algebraic manipulations can
be used to show that
\begin{equation}
\sigma^{\cS}_{|Y_i} = \frac{\Tr_\cF \{Y_i \sigma^{\cS\cF}
Y_i\}}{p(Y_i)} = \sum_j p(X_j | Y_i) \sigma^\cS_{|X_j}.
\end{equation}
We see that $\sigma^{\cS}_{|Y_i}$ is a convex combination of the
$\sigma^{\cS}_{|X_j}$. (This is no coincidence: it follows from
Corollary~\ref{corollary:effect}, \ref{lemma:commute}, and the
existence of information about $\bA$, and hence about $\bX$, in the
rest of the environment $\overline\cF$.) Hence, the inequality of
Eq.~(\ref{eq:measure_opt}) follows from the convexity of entropy,
c.f. Proposition~\ref{prop:convexity}. Similarly, when $\bY$ is a
fixed measurement on $\cF$, maximizing $I(\bB:\bY)$ amounts to
minimizing $H(\bY|\bB)$. We can consider the state of $\cF$ following
a measurement of $\bB$ on $\cS$
\begin{equation}
\sigma^{\cF}_{|B_i} = \frac{\Tr_\cS \{B_i \sigma^{\cS\cF}
b_i\}}{p(B_i)} = \sum_j p(A_j | B_i) \sigma^\cF_{|A_j},
\end{equation}
a convex combination of the states of $\cF$ conditioned on a
measurement of $\bA$ on $\cS$. Again, the inequality
Eq.~(\ref{eq:observable_opt}) follows from convexity of entropy,
c.f. Proposition~\ref{prop:convexity}.

By combining Eq.~(\ref{eq:measure_opt}) with the fact that measuring
$\bX$ perfectly emulates the direct measurement $\bA$ on $\cS$, we get
the following equality:
\begin{equation}\label{eq:max_info}
\hat I_\cF(\bB) = I(\bB:\bA).
\end{equation}
The information about $\bB$ accessible from the fragment $\cF$ is
inherently limited by its correlation with the maximally refined
observable $\bA$. The only assumption required to arrive at this
important equality is that correlations with both the observed and the
unobserved parts of the environment impose $\gamma^{\cF}_{ij} =
\gamma^{\overline \cF}_{ij} = \delta_{ij}$. This requirement is,
according to Corollary~\ref{corollary:induced}, equivalent to saying
that both $\cF$ and $\overline \cF$ contain a perfect copy of the
information about $\bA$.

The consequences of Equation~(\ref{eq:max_info}) can be better
appreciated once we recognize that it allows to evaluate the
redundancy for any system observable straightforwardly. When the value
of $\bB$ can be deduced from knowledge of the value of $\bA$, then
each fragment of $\cE$ containing information about $\bA$ will
inevitably contain information about $\bB$. Formally,
\begin{equation}
I(\bB:\bA) \geq (1-\delta) \hat I_\cE(\bB) \Rightarrow R_\delta(\bB) \geq
R_{\delta =0}(\bA).
\label{eq:min_redundancy}
\end{equation}
On the other hand, when the value of $\bB$ {\em cannot} be deduced
from knowledge of $\bA$, $\bB$ cannot be redundant:
\begin{equation}
I(\bB:\bA) < (1-\delta) \hat I_\cE(\bB) \Rightarrow R_\delta(\bB) =1.
\label{eq:max_redundancy}
\end{equation}
This is because Eq.~(\ref{eq:max_info}) holds as long as $\bar\cF$
contains information about $\bA$. To have $\hat I_{\cF}(\bB) \geq \hat
I_{\cE}(\bB)$, forces to take $\bar\cF$ small enough so that it does
not contain much information about $\bA$. By virtue of
Eq.~(\ref{eq:measure_opt}), $\bar\cF$ also contains little information
about $\bB$, implying $R_\delta(\bB) = 1$.

Note that these results can also be derived with the help of the data
processing inequality (see Proposition~\ref{prop:dpi}). In effect, the
update rule of Eq.\@~(\ref{eq:state_update}) shows that the sequence
of measurements $\bB$, $\bA$, $\bX$, $\bY$ forms a Markov chain ($\bB$
and $\bY$ are arbitrary): the joint probability of the measurement
results of $\bB$, $\bA$, $\bX$, $\bY$ satisfies
\begin{equation}
p(Y_i, X_j, A_k, B_l) = p(Y_i|X_j)p(X_j|A_k)p(A_k|B_l)p(B_l),
\end{equation}
yielding Eqs.~(\ref{eq:measure_opt},\ref{eq:observable_opt})
directly.  This alternative derivation provides a very clear
interpretation of our previous result. Trying to gather information
about $\bB$ (instead of $\bA$) with the indirect measurement $\bY$
(instead of $\bX$) can be viewed as the addition of noise over the
perfect communication channel that allows the transmission of
information about $\bA$ in the environment.

Equations (\ref{eq:min_redundancy}-\ref{eq:max_redundancy}) extend Theorem~\ref{th:uniqueness} to
imperfect redundant imprints (i.e. finite $\delta$). Only observables
``close'' to the maximally refined redundant observable $\bA$ --- where
closeness is measured with the help of mutual information --- can
leave a redundant (even imperfect) imprint in their
environment. Moreover, it confirms the behavior found in the numerical
study presented in \cite{OPZ04b}.

\subsection{Imperfect correlations} \label{sec:model_imperfect}

The previous section analyzed the consequences of perfect correlations
between $\cS$ and $\cF$ in terms of optimal measurement
strategies. However, perfect correlations are rarely found in
Nature. Even for our simple model, perfect correlations can be assumed only in
an asymptotic limit, as $N$ tends toward infinity, or by a careful
tuning of the interaction time and strength. Hence, it is important to
understand what happens when the conditions are not perfect. Here, we
show that nearly perfect correlations --- $\gamma^\cF$ and
$\gamma^{\overline \cF}$ sufficiently small --- are enough to ensure
the validity of the results established above, up to small correction
terms.

The technique we use is inspired by perturbation theory. It relies on
the construction of a ``perfectly correlated'' state $\sigma^{\cS\cF}$
of the form Eq.~(\ref{eq:classical_state}) which is ``close'' to the
actual state $\rho^{\cS\cF}$ of Eq.~(\ref{eq:state_real}) generated by
the dynamics. Then, using various bounds on entropies, we will
conclude that all the information-theoretic quantities extracted from
the ideal $\sigma^{\cS\cF}$ are approximately equal to those
extracted from the actual $\rho^{\cS\cF}$.  We will also examine the
regime of validity of this approximation.

Let us define
\begin{equation}
\sigma^{\cS\cF} = \sum_i |\alpha_i|^2 \kb i i \otimes \kb
{\Psi^\cF_i}{\Psi^\cF_i},
\end{equation}
where the $\ket{\Psi^\cF_i}$'s are obtained by applying the
Gram-Schmidt orthonormalisation procedure to the states $\ket
{\Phi^\cF_i}$'s (see the appendix for the details of this
construction). With this definition, we will show that for any two
observables $\bB$ and $\bY$, when $\gamma^\cF =
\max_{ij}{|\gamma^\cF_{ij}|}$ and $\gamma^{\overline \cF} =
\max_{ij}{|\gamma^{\overline \cF}_{ij}|}$ are small,
\begin{equation}
I_\rho(\bB:\bY) \approx I_\sigma(\bB:\bY).\label{eq:I_approx}
\end{equation}
Above, the subscripts $\rho$ and $\sigma$ refer to the state, either
the actual $\rho^{\cS\cF}$ or the perfect $\sigma^{\cS\cF}$, used to derive the
probabilities of the measurement outcomes, that in turn are used to
quantify information (this shorthand notation will be used in the rest
of this section). Therefore, when Eq.\@~(\ref{eq:I_approx}) holds, all
the conclusions derived from the perfect correlation case remain
approximately true.

Equation~(\ref{eq:I_approx}) is a consequence of simple inequalities
that give an upper bound on the difference $|I_\rho(\bB:\bY) -
I_\sigma(\bB, \bY)|$. First, Cauchy-Schwartz inequality
(Proposition~\ref{prop:CS}) gives $\Tr\, |\rho^{\cS\cF} -
\sigma^{\cS\cF} | \leq \sqrt{d^{\cS}} \|\rho^{\cS\cF} -
\sigma^{\cS\cF}\|_2$. Second, by definition of the trace distance
between two density matrices (Definition~\ref{def:trace_distance}),
$\sum_{ij} |p_\rho(B_i, Y_j) - p_\sigma(B_i, Y_j)| \leq \Tr\,
|\rho^{\cS\cF} - \sigma^{\cS\cF}|$. Finally, by combining these
results with Lemma~\ref{lemma:GS} and Fanne's inequality
(Proposition~\ref{prop:fanne}) applied separately to each of the three
entropies involved in $I(\bB, \bY) = H(\bB) + H(\bY) - H(\bB, \bY)$,
we find
\begin{eqnarray}
&& |I_\rho(\bB:\bY) - I_\sigma(\bB:\bY)| \\ && \quad \leq
-3f(\gamma^\cF, \gamma^{\overline\cF}) \log (f(\gamma^\cF,
\gamma^{\overline\cF}) / d^\cS) \nonumber \\ && \qquad +
O((\gamma^\cF)^{3/4} + \gamma^{\overline \cF}), \nonumber
\end{eqnarray}
with $f(\gamma^\cF, \gamma^{\overline \cF}) = \sqrt{d^\cS
\{2(d^\cS-1)(\gamma^\cF)^2 + (\gamma^{\overline \cF})^2\}}$. Thus,
this difference tends to zero when $\gamma^\cF$ and
$\gamma^{\overline\cF}$ tend to zero.

To gain further insight into the regime of validity of
Eq.\@~(\ref{eq:I_approx}), recall that $\gamma^\cF$ and
$\gamma^{\overline \cF}$ typically decrease exponentially with the
number of environmental subsystems that have effectively interacted
with $\cS$. Therefore, for equation Eq.\@~(\ref{eq:I_approx}) to hold
within constant accuracy, it is sufficient that each fragment $\cF$
and $\overline \cF$ contains a number of subsystems $\cE_k$ scaling as
$\log d^\cS$ (or as a polynomial in $\log d^\cS$). As discussed at the
end of Section~\ref{sec:defs}, $d^\cS$ is the effective dimension of
the {\em probed} degree of freedom of $\cS$, i.e. the number of
distinguishable outcomes of an hypothetical measurement $\bB$. Clearly
then, $\log d^\cS$ is generally much smaller than the size of the
environment.  Therefore, the conclusions drawn from the perfect
correlation study remain essentially unchanged: $\bA$ is the
observable of the system that leaves the strongest imprint in
fragments of the environment; the optimal environmental measurement to
learn about any system observable is the one that reveals information
about $\bA$; and finally failing to interrogate $\cF$ about $\bA$ has
the same effect as introducing noise in the measurement results. These
conclusions hold to within an accuracy of roughly $\log d^\cS /N$,
which is enough in most situations involving macroscopic systems.

\section{Discussion: Open questions and connections}
\label{sec:discuss_model}

Environment as a witness as well as quantum Darwinism --- the dynamics
responsible for the redundant imprinting of certain states that leads
to their objectivity --- are based on the same dynamical paradigm that
is used in the study of decoherence and einselection. The
to-be-classical ``object of interest'' (the system $\cS$ or the
apparatus ${\cal A}$) is immersed in the environment. What is now
different --- and this is a dramatic departure from the usual view of
what matters --- is the focus.  Instead of analyzing either the state
of $\cS$ {\it per se}, or even the fate of the correlations between
$\cS$ and ${\cA}$, attention shifts to the information available to
the observer in {\em fragments} of $\cE$.

This shift of focus has been precipitated by the realization
\cite{Zur93b, Zur98a, Zur00a} that observers only rarely (if ever)
acquire information by direct interaction with ``systems of
interest'': rather, we use $\cE$ as a communication channel.
The message --- the state of the object of interest --- is imprinted
on $\cE$ in multiple copies. However, the {\it no-cloning theorem}
implies that arbitrary unknown states cannot be ``advertised''
throughout the environment in this fashion. Hence, as one can
expect (and as our analysis indeed shows), selection of a preferred 
observable is a prerequisite for this redundant imprinting.

The basic intuition --- the implication of the information-theoretic
redundancy of the record for the emergence of the classical --- was
noted already some time ago \cite{Zur82a, Zur83a, Zur00a}. It was
refined into a more precise measure of redundancy --- into
$R_{\delta}$ we employed here --- only recently \cite{OPZ04b}.  Using
this $\delta$-redundancy we have demonstrated --- either under very
specific assumptions of \cite{OPZ04b}, or in a still rather specific,
but somewhat more general setting of this paper --- that the familiar
pointer observables are easiest to find indirectly, from the records
imprinted on the fragments of the environment.

The aim of this section is to examine assumptions that went into our
discussion and, by doing so, to explore the range of validity of our
conclusions.  The results obtained in Section~\ref{sec:theory} did not
require any assumption about the physical model, but only apply to the
case of perfect correlations. Thus, the model studied in
Section~\ref{sec:model} is a convenient focus of attention. To
establish uniqueness of a set of commuting preferred observables that
is easiest to infer from $\cE$ we have assumed that:
\begin{enumerate}
\item The initial state is a pure product state.
\item The system has no self-Hamiltonian.
\item Every environmental subsystem couples to the same system
observable.
\item The environment has no Hamiltonian.
\end{enumerate}
To find out which of these assumptions can be relaxed, and the extent
to which they are responsible for the conclusions we have reached will
eventually require investigation of other, more realistic, or at least
``differently oversimplified'' models. For ``standard'' decoherence, a
similarly idealized model was put forward over two decades ago, but
the investigation of various related and unrelated models of
decoherence and einselection is still an ongoing activity, often
yielding new insights. The same can be expected of quantum
Darwinism. Different models may also require different mathematical
tools. One such investigation --- using von Neumann entropy and the
symmetric Equation~(\ref{eq:mutual_sym}) for mutual information --- is currently under way
\cite{BKZ04a} and points toward compatible conclusions.\footnote{The
two formulae for the mutual information, Eqs.~(\ref{def:mutual}) and (\ref{eq:mutual_sym}), cease to
coincide when the von Neumann entropy of the two subsystems and the
relevant joint and conditional entropies are used on their left hand
sides. The difference between the symmetric and asymmetric mutual
information is the {\it discord} \cite{Zur00a, OZ02a, Zur03a}. Discord
is a measure of the {\it quantumness} of the correlation.  It can be
used to show that even separable correlated states of two systems need
not be classical. Discord is suppressed by decoherence. So, when a
small fraction of the environment is interrogated by the observer, its
correlations with $\cS$ tend to be effectively classical. It follows
that --- in that case -- estimates of redundancy based on either the
mutual information defined by Equation~(\ref{def:mutual}) or by Eq.~(\ref{eq:mutual_sym}) are essentially
identical.}  All we can offer here is a brief (and possibly premature)
discussion of the role and importance of various assumptions,
including these listed above.

{\bf Assumption 1} is probably the most unrealistic: in practice
composite environments are rarely in a pure product state. Luckily,
perfect purity {\it per se} is not essential to rely on the
environment as a trustworthy witness. Having enough information about
$\cE$ will, however --- and in contrast to decoherence and
einselection --- prove to be indispensable.

Indeed, it is often convenient to study decoherence assuming a very
mixed state of the environment (i.e., thermal equilibrium or even a
perfect mixture) when deriving master equations used to implement
predictability sieve. Consequences of different initial mixtures of
$\cE$ are, generally, somewhat different time dependences, and minor
changes in the structure of the master equation (which often becomes
less Markovian and less tractable when the environment is further away
from a convincing thermal state), but the key qualitative conclusions
that characterize decoherence and einselection --- the localized
nature of the preferred states and the fact that relaxation can be
much slower than decoherence --- seem to be usually unaffected.

By contrast, the degree of ignorance about the fragments of the
environment plays an essential role in the study of the environment
as a witness. The reader may be surprised by this. After all, the results
of Section~\ref{sec:theory} and \ref{sec:model} were derived by an
information-theoretic analysis of a rather general scenario, so ---
given the assumptions --- conclusions should be independent of the
initial state of $\cE$. In particular, redundancy --- when present ---
remains a highly selective criterion.  However, and this is the crux
of the matter, the starting point of our investigation --- the assumption 
that fragments of the environment contain information about the system 
--- depends strongly on the initial state of $\cE$.

When the environment is initially mixed, it will ``know less'' about
the system.  The {\it change} of the state of the environment after it
has interacted with the system provides the evidence about the state
of $\cS$. But when either all of $\cE$ or some subsystems of it are
totally mixed, they cannot be altered by the conditional dynamics that
causes decoherence. And even partial mixing will make it more
difficult for the environment to serve as a wittness. To see why, let
the initial state of the environment be an arbitrary non-entangled state:
\begin{equation}
\rho^\cE = \sum_\ell p_\ell \bigotimes_{\cE_k}
\kb{\phi^{\ell\cE_k}}{\phi^{\ell\cE_k}}.
\label{eq:general_state}
\end{equation}
In particular, it could be a product of mixed states $\rho^\cE =
\rho^{\cE_1}\otimes\rho^{\cE_2}\otimes\ldots\rho^{\cE_N}$, such as the
thermal state of non-interacting environmental subsystems. By
linearity of the Schr\"odinger equation, the joint state of $\cS$ and
a fragment $\cF$ of $\cE$ after an interaction time $t$ will be
\begin{equation}
\rho^{\cS\cF} = \sum_\ell p_\ell \sum_{ij} \alpha_i\alpha_j^* \kb ij
\bigotimes_{\cE_k} \kb{\phi_i^{\ell\cE_k}}{\phi_j^{\ell\cE_k}} \times
\gamma_{ij}^{\ell\overline\cF}
\end{equation}
where $\ket{\phi_i^{\ell\cE_k}}$ and $\gamma_{ij}^{\ell\overline\cF}$
are defined as in Section~\ref{sec:model_dynamics}.

In the previous sections, we have studied the kinds of correlations
that arise given any of the components $\bigotimes_{\cE_k}
\ket{\phi^{\ell\cE_k}}$ of this mixture. It follows from the convexity
of entropy (c.f. Proposition~\ref{prop:convexity}) that the
information contained in this mixture will be strictly less than the
average information contained in the individual components of the
mixture. As we have noted above, mixing the state of the environmental
subsystems decreases their information storage capacity; when the
state of the environment is perfectly mixed, it will simply be
impossible to learn anything from it. Yet, the usual decoherence still
leads to einselection --- there will still be a pointer basis (defined
e.g. through the predictability sieve).

When the initial state of $\cE$ is not completely mixed, it is usually
possible to extract from it some information about the system. In
fact, given the assumption of redundancy, the qualitative conclusions
of the previous section will still hold: the information about any
system observable available in fragments of $\cE$ is intrinsically
limited by the information about it obtained through a direct
measurement of the maximally refined redundantly imprinted observable
$\bA$; Eq.~(\ref{eq:max_info}) becomes the inequality $\hat I_\cF(\bB)
\leq I(\bB :\bA)$. Therefore, we again anticipate the conclusion that
only observables ``close'' to $\bA$ can get redundantly imprinted in $\cE$. 

Mixed initial states of $\cE$ will decrease channel capacity of the
environment.  Moreover, modes of the environment that can be imprinted
with information about different aspects of $\cS$ will be in general
mixed to a different degree. This may be reflected in the selective
coarse-graining with which the observer perceives the system.  This
raises an interesting possibility: observables that are selected by
the predictability sieve need not coincide with the (typically coarse
- grained) observables that are easiest to find out about from
$\cE$. For example, this may happen when the modes that dominate
decoherence (and hence einselection) are so mixed that they cannot
serve as a useful communication channel, and the only modes that are
sufficiently pure to be useful for this purpose monitor some other
observable. An observer will then find out about that other
coarse-grained observable, but with the resolution limited by
indeterminacy: after all, pointer observable is still einselected by
the environmental monitoring! Presence or absence of this effect is
related to the assumption of the ``typical environment fragment'' we
have noted in \cite{OPZ04b}. In other words, when the fragments of $\cE$ that the
observer can use to extract information about $\cS$ contain or reveal
data in a selective manner, environment acts as a witness with a
selective memory (or a partial amnesia). Exactly how mixed can $\cE$ be to be
still useful as a witness is of course the key question. We leave it
as a subject for further research.

When the initial state of the {\em system} is mixed, but the
environment is sufficiently pure and large so that significant
redundant imprinting can arise from their interaction, environment can
become a useful and trustworthy witness, and our approach should go
through essentially unimpeded.  Similarly, preexisting correlation
with $\cE$ need not undermine our conclusions. In effect, it will
typically mean that the environment was gathering evidence about the
system in the past (although one can certainly imagine pathological
preexisting correlations that get undone by the subsequent
interaction).

To sum up, while the assumption of the initial product state has
simplified our analysis, it can be relaxed to some extent without
undermining our basic conclusions.  What is non-negotiable is the 
demand that the final state allows one to discern evidence about $\mathcal S$ in the fragments of $\mathcal E$. One can already anticipate that this
demand will also put a constraint on the entanglement between the
fragments of $\cE$, as we shall note in the discussion of the
assumption 4 below.

{\bf Assumption 2} precludes non-trivial evolution of the system. 
Relaxing it has been studied in the context of decoherence and
einselection, and --- as we have done above --- we shall gain insight
into its role in quantum Darwinism by recounting the implications of
evolution for einselection, and analyzing their significance for the
role of the environment as a witness.

In the studies of einselection the relative strength of the
self-Hamiltonian $\bH^\cS$ and the interaction Hamiltonian $\bH^{\cS \cE}$ 
is an important ingredient that decides the course and
effect of einselection. When the self-Hamiltonian is negligible or
when $\bH^\cS$ and $\bH^{\cS\cE}$ commute, the environment
monitors static states of the system --- pointer states selected
solely by the interaction \cite{Zur81a, Zur82a}. In a sense, environment
contains a record of a very uneventful history --- things stay the
same forever.

When the effect of $\bH^\cS$ is no longer negligible, these
histories become more interesting. One may enter, for example, 
the ``partial Zeno effect" regime where the evolution of the state of the system 
induced by the self-Hamiltonian is impeded (but not completely 
suppressed) by the environmental
monitoring mediated by $\bH^{\cS\cE}$ (see, e.g., \cite{GJK+96a}). 
Models of such situations have been analyzed using rather different 
approach of quantum trajectories \cite{Car93a} that shares with quantum 
Darwinism the focus on the information gained indirectly from $\cE$. There 
the environment is usually assumed to consist of photons, and the system 
is often a two-level ``atom''. Pointer states still emerge and turn out to be 
the easiest states to infer from a fraction of such $\cE$ \cite{DDZ01a}.  
The basis ambiguity may reappear only if the whole $\cE$ could be captured.  
This ability to rely on a fraction of $\cE$ allows for the possibility of consensus
between multiple observers \cite{DDZ04a}. 

As the coupling to the environment becomes weak the problem may 
become more tractable again, especially in two rather different limits. 
When the environment is slow, with the high-frequency cutoff that is small 
compared to the level spacing of the spectrum of $\bH^\cS$, einselection
enters its quantum limit \cite{PZ99b}. Energy eigenstates are
imprinted on the environment more or less regardless of the specific
form of the coupling.  Histories are again becoming rather uneventful,
although for reasons quite different from these we have mentioned
above in the negligible $\bH^\cS$ case.

By contrast, when the environment is ``broadband'', with all the
frequencies from very low to very high present, and the system is (at
least approximately) linear and coupled to $\cE$ through its position,
coherent states are the most predictable, and hence, pointer. 
Weak coupling does not preclude redundancy of the records. In this 
underdamped limit one can recover approximately reversible Liouvillian  
dynamics with localized states (i.e., approximately reversible classical
trajectories that follow Newton's laws emerge; see
\cite{PZ01a, Zur03a}, and references therein). When the system is
linear (e.g., harmonic oscillator) preferred pointer states turn
out  to be Gaussian \cite{Zur93b, ZHP93a}, and one may anticipate
that sequences  of the most redundant ``environmental records'' will
correlate with  classical trajectories.

The quick summary of different possibilities above is meant to be
suggestive rather than exhaustive, and is definitely not
conclusive. What one would like to recover using environment as a
witness is the notion of objective histories defined through
sequences of time-ordered redundant records. Such objective  histories
could consist of sequences of approximate instantaneous pointer
states. They were conjectured to exist in the limit of very efficient
and  frequent broad-band monitoring by the environment in
\cite{Zur03a}.  Histories deduced from redundant records would be
objective, and may be only  approximate. It will be interesting to
investigate their relation to consistent  histories \cite{Gri84a,
GH90a, Omn92a, Omn99a, Gri02a}, especially when consistency is assured
through the ``strong decoherence'' condition introduced in \cite{GH97b}
(see also \cite{Hal99b}) that invokes existence of a perfect
environmental  record.

{\bf Assumption 3} assures that there is a preferred observable of
$\cS$ that is recorded by $\cE$, and that it is singled out by the
interaction Hamiltonian.  This is a strong assumption, but often a
reasonably realistic one: typical couplings between the systems (and,
hence, between $\cS$ and $\cE$) tend to depend on the distance
(i.e., relative position) between them. As a consequence --- as was 
already pointed out in the context of einselection \cite{Zur82a, Zur91a} 
--- pointer states tend to be localized in position.

One can of course imagine situations where this is not the case. For
instance, there may be several physically distinct environments, each
attempting to  monitor a different observable of $\cS$. One would
expect that a dominant  coupling would then impose its selection of
the preferred observable, and the effect of the others would be
perceived as noise.

This last remark leads to interesting questions about the correspondence between the approximate pointer observable that is selected by the predictability sieve as a result of the interaction of $\mathcal S$ with $\mathcal E$, and the states of the system that are easiest to infer for an observer who has access to a single subsystem of the environment. Clearly, in full generality this case is not covered by our idealized
assumptions earlier in the paper. For example, Gaussian pointer states are 
often selected by the predictability sieve, and yet they are not orthogonal 
and form an overcomplete basis, so some of the steps we have used
in our proofs may not go through. Nevertheless, one may expect that
--- as was the case with decoherence --- results that are exact for
perfect pointer states (that are left untouched by decoherence)
are still approximately valid for approximate pointer states (that are least
touched by decoherence). Thus, even when the environment fragments 
that are the most useful channel to the observer interested in $\cE$ are not 
a typical sample of the environments responsible for decoherence, one may 
still guess that the states of $\cS$ that can be inferred from them will represent 
an appropriately coarse-grained versions of the pointer states. Given the role Gaussian states play in the discussion of various models of decoherence, the extent to which Gaussian states coincide with the states that are easiest to infer is of obvious interest, and a likely area of future studies, but beyond the scope of this paper (see however \cite{DDZ05a}). Along the same line of thought, when the environment fragments that are the most useful to the observer interested in $\mathcal S$ are not a typical sample of the environments responsible for decoherence, one may still guess that the states of $\mathcal S$ that can be inferred from them will represent an appropriately coarse-grained versions of the pointer states.

Given
the role Gaussians play in discussion of various models of decoherence, 
the extent to which Gaussians that are most predictable coincide with the 
states that are easiest to infer is of obvious interest, and a likely area of future 
studies, but beyond the scope of this paper.

{\bf Assumption 4} assures that the environment does not evolve (and,
therefore, does not complicate or even obliterate the record it has
made of $\cS$). More careful study of the consequences of relaxing this
assumption can be found elsewhere \cite{BKZ}.
Here we offer only a brief discussion of the basic possibilities.

One can distinguish three components of this requirement: fragments of
the environment could evolve separately, they could interact with each
other, or could interact with more distant ``second order''
environments. 

Let us first note that --- in models where pointer observable commutes 
with the system-environment interaction Hamiltonian 
(e.g. ${\bf H}^{\cS\cE} = \sum_{\cE_k \in \cE} \bA \otimes \bZ^{\cE_k}$,
Eq.~(\ref{eq:coupling})) self-Hamiltonians of the 
individual fragments of $\cE$ obviously cannot change the fact that
$[{\bf H}^{\cS\cE} ,\bA]=0$, so the selection of the pointer basis is not
affected by the dynamics of the environment. However, even in this 
simple case the presence of such self-Hamiltonians can influence the
rate of acquisition of information by $\cE$ and, hence, redundancy
\cite{BKZ}. We also note that in more 
realistic cases when the system has a self-Hamiltonian that does not commute
with ${\bf H}^{\cS\cE}$ the pointer observable is approximate, and even
its choice can be influenced by how the rate of information acquisition by
$\cE$ compares with the dynamical timescales in $\cS$.

The interaction between the subsystems of $\cE$ may be a
significant complication. It will typically lead to entanglement
between the fragments of $\cE$, and could make their individual states
mixed. Thus, --- as we have already pointed out in discussion of
assumption 1 --- even though the environment as a whole may still
contain redundant imprint  of the state of the system, the relevant
global observables could become  effectively inaccessible to the
observers who can sample $\cE$ only through  local measurements on its
fragments.

Last but not least, $\cE$ could be immersed in its own environment
$\cE'$. Effect of this second-order environment will depend on the
nature of their interaction. It could of course erase the records of
$\cS$ in $\cE$, but it is also conceivable that $\cE'$ could monitor
the record-keeping observables of the fragments of $\cE$, which would
increase the redundancy \cite{Zur03a}.

These are very interesting complications, very much worthy of further
study.  We note, however, that human observers tend to rely on photons
for the vast majority of data about ``systems of interest''. And
photons, in effect, satisfy our assumption 4. This seems to us to be
no accident.

It is clear from the above discussion that quantum Darwinism requires
stronger assumptions than decoherence. This is not surprising: it is
easier to use any system (including the environment) as a disposal, 
to get rid of  unwanted information, than as a communication channel. 
Furthermore, existence of the pointer states is a necessary but not a 
{\em sufficient} condition for their redundant imprinting on $\cE$. 
Nevertheless, even though quantum Darwinism relies on stronger assumptions 
than einselection implemented e.g. through the predictability sieve, 
it is also  clear from our discussion that
these more restrictive assumptions  are met in the real world often
enough to employ environment as a reliable  witness.

\section{Summary and Conclusion}
\label{sec:conclusion}

Observers use environment as a witness. This  has obvious 
and profound implications for the interpretation of quantum theory.
Nevertheless, it is not our intent here to go beyond the comments 
we have made throughout and to delve into a more extensive 
discussion of the implications of quantum Darwinism for the  
interpretation of quantum theory. In part, this is because some 
of the interpretational issues have been noted elsewhere,
on the occasion of discussions of the role of redundancy in the
{\em existential interpretation} (see \cite{Zur93b, Zur98a, Zur03a, Zur03b},
and references therein). The basic  idea --- that the classical domain
can arise from  within a quantum Universe,  and that it consists of
states that can be found out without being destroyed,  and, hence, in
that operational sense exist independently of the observer ---  is
obviously very much assisted by employing environment as a witness.

Quantum Darwinism uses the same model of the information transfer that
was introduced to study decoherence and einselection, but asks a
different question: instead of focusing --- as does decoherence --- on
how the information is lost from the system, it analyzes how the
information is gained and stored in the environment. In decoherence,
the role of the environment is limited to hiding the underlying
quantumness (view especially emphasized in discussions of quantum
computation). Einselection recognizes that this information loss is
selective, and, thus, that the environment has a capacity to single
out preferred pointer states that are best at surviving immersion in
$\cE$. Moreover, they are chosen in the process reminiscent of a
measurement --- $\cE$ acts as an apparatus that (pre-)measures the
pointer observable of the system of interest.

This much was known. In particular, the monitoring role of $\cE$ was
recognized early on \cite{Zur81a}. Quantum Darwinism answers to the
next logical question: since the environment interacts with the
to-be-classical observable of the system of interest as would an
apparatus, can it be used as an apparatus?  To address it we have had
to develop a measure of the objectivity of the records deposited in
the environment. Redundancy (briefly considered \cite{Zur83a}, see
also \cite{Zur82a} in a similar context as a measure of amplification)
was refined \cite{OPZ04b} and used to prove that --- given reasonable
assumptions --- there is a unique set of commuting observables that
are easiest to find out from fragments of $\cE$, and that these
redundantly recorded observables are indeed the familiar pointer
observables.

As a consequence, observers monitoring fragments of the environment
will be able to reach compatible conclusions about the value of
redundantly imprinted observables of the system, i.e. about the {\em
objective} properties of the system. In this operational
sense, quantum Darwinism provides a satisfying explanation for the
emergence of the objective classical world we perceive from the
underlying quantum substrate.

We have illustrated quantum Darwinism on a dynamical model that is
easy to analyze from an information-theoretic perspective, and yet is
inspired by a photon environment scattering on an object: all the
environmental subsystems couple to the object with the same
Hamiltonian and do not interact with each other. This model enabled us
to push our analysis further, as we could explicitly determine the
optimal environmental measurements to learn about the properties of
the system. It also demonstrated that quantum Darwinism tolerates
reasonable departures from the ``ideal'' assumptions used to derive
our main results in Section~\ref{sec:theory}. In this respect, it
corroborates recent studies \cite{Zur03a,OPZ04b} and earlier
conjectures \cite{Zur83a, Zur00a} that noted the role of redundancy in
assuring {\em resilience} of certain states --- an essential feature
of the classical domain. Hence, redundancy is a robust and selective
criterion for determining objective properties of open quantum
systems.

While redundantly imprinted observables are dynamically stable --- as
defined by einselection --- the reverse is not necessarily true:
quantum Darwinism requires stronger assumptions, especially regarding
structure, size, and the initial state of $\cE$. For example, a single
photon can be enough to decohere an object in superposition of two
distinct positions. However, it takes a macroscopic number of them to
redundantly broadcast the position of this object throughout the
environment. While the entropy production of the system --- a
signature of decoherence --- rapidly saturates, redundancy can then
continue to grow with time. This illustrates how the increase of the
redundancy of the record in the environment captures the fact that the
system is still continuously ``under observation'', and that
information about the pointer observable is getting amplified up to
the macroscopic level. In this sense, quantum Darwinism can be
motivated using Bohr's original ideas \cite{Boh28a} about the role of
{\em amplification} in the transition from quantum to classical.

\section{Acknowledgments}
We would like to thank Robin Blume-Kohout for extensive and enjoyable
discussions.
This research was supported in part by NSA and ARDA. H.O. receives
financial support from ACI S\'ecurit\'e Informatique, and D.P. from
Canada's NSERC and Qu\'ebec's NATEQ.

\section{Appendix}

\begin{definition}[Trace distance]\label{def:trace_distance}  
The trace distance between two density operators $\rho$, $\sigma$ is
defined by $D(\rho, \sigma) = \Tr\, |\rho - \sigma|$. When $\rho$
and $\sigma$ commute, the trace distance is equal to the
$L_1$-distance between the classical probability distributions defined
by their eigenvalues. Alternatively, the trace distance is equal to
$2\max_P \Tr\, P(\rho - \sigma)$ where the maximum is taken over
positive operators $P \leq \Id$
\end{definition}

\begin{definition}[Euclidean distance]
The Euclidean distance between a pair of density operators $\rho$,
$\sigma$ is defined by $\|\rho-\sigma\|_2 = \sqrt{\Tr\left\{(\rho -
\sigma)^2\right\}}$.
\end{definition}

\begin{proposition}[Cauchy-Schwartz inequality] \label{prop:CS}
$D(\rho,\sigma) \leq \sqrt{d} \|\rho-\sigma\|_2$.
\end{proposition}

\begin{proposition}
Let $p_i = \Tr\{A_i\rho\}$ and $q_i = \Tr\{A_i\sigma\}$ be the
probabilities of the outcomes of a measurement $\{A_i\}$ for states
$\rho$ and $\sigma$ respectively. Then, $D(p,q) \leq D(\rho,\sigma)$.
\end{proposition}

\begin{proposition}[Fanne's inequality]\label{prop:fanne} 
Let $p_i$ and $q_i$ with $i=1,2,\ldots , d$ be two classical
probability distributions such that $D(p,q) \leq 1/e$. Then $|H(p) -
H(q)| \leq D(p, q) \log d + \eta (D(p, q))$, with $\eta(x) = -x \log
x$.
\end{proposition}

\begin{proposition}[Convexity of entropy]\label{prop:convexity}
Let $P(X_i|Y_j)$ be a set of conditional probability distribution for
the random variable $X$, and $P(Y_j)$ be the probability distribution
for the random variable $Y$. Then
\begin{eqnarray}
&& \sum_j \left[P(Y_j) \left( -\sum_i P(X_i|Y_j) \ln P(X_i|Y_j) \right)\right]
\nonumber \\ && \quad \leq -\sum_i P(X_i) \ln P(X_i)
\end{eqnarray}
where $P(X_i) = \sum_j P(X_i|Y_j)P(Y_j)$.
\end{proposition}

\begin{proposition}[Data processing inequality]\label{prop:dpi}
Suppose $X \rightarrow Y \rightarrow Z$ is a Markov chain, then $H(X)
\geq I(X:Y) \geq I(X:Z)$.
\end{proposition}

\begin{lemma}\label{lemma:GS}
For any matrix $\rho = \sum_{ij} \alpha_i \alpha_j^* \kb{i}{j} \otimes
\kb{\Phi^\cF_i}{\Phi^\cF_j} \gamma_{ij}^{\overline \cF}$, with
$\alpha_i \neq 0$, and such that,
\begin{enumerate}
  \item $\bk i j = \delta_{ij}$
  \item $\ket {\Phi^\cF_i}$ are linearly independent;
  \item $|\gamma^{\overline \cF}_{ij}| \leq \gamma^{\overline \cF}$
  for $i\neq j$, and $\gamma^{\overline \cF}_{ii} = 1$ for all $i$;
  \item $|\bk{\Phi^\cF_i}{\Phi^\cF_j}| = |\gamma^\cF_{ij}| \leq
  \gamma^\cF$ for $i \neq j$;
\end{enumerate}
there exist a matrix $\sigma = \sum_{i} |\alpha_i|^2 \kb{i}{i}
\otimes \kb{\Psi^\cF_i}{\Psi^\cF_i}$, satisfying
\begin{enumerate}
  \item $\bk{\Psi^\cF_i}{\Psi^\cF_j} = \delta_{ij}$;
  \item $\| \rho - \sigma \|_2 \leq \sqrt{2(d^\cS-1)} \gamma^\cF +
  \gamma^{\overline \cF} + O((\gamma^\cF)^{3/2} + (\gamma^{\overline
  \cF})^2)$.
\end{enumerate}
where $d^\cS$ is the number of coefficients $\alpha_i$.
\end{lemma}

\begin{proof}
The proof of this lemma relies on a careful analysis of Gram-Schmidt
orthonormalisation procedure applied to the states $\ket{\Phi^\cF_i}$.

We define the new set of orthogonal states $\ket{\Psi_i^\cF}$ by the
following equations:
\begin{equation}
\ket{\Psi^\cF_i} = \frac{1}{N_i}\left(\ket{\Phi^\cF_i} - \sum_{j<i}
\bk{\Psi^\cF_j}{\Phi^\cF_i}\ket{\Psi^\cF_j}\right), \label{eq:GS}
\end{equation}
where $N_i$ is a normalization factor.

More precisely, we prove by induction on $i$ that the following
properties hold:
\begin{equation}
\left | \bk {\Phi^\cF_j}{\Psi^\cF_i}\right | \left \{
\begin{array}{ll} \leq 2 \gamma^\cF + O((\gamma^\cF)^2), & j < i \\
\geq 1 - \frac{i-1}{2} (\gamma^\cF)^2 + O((\gamma^\cF)^3), & j = i \\
\leq \gamma^\cF + O((\gamma^\cF)^2), & j > i
\end{array}
\right.\label{eq:rec}
\end{equation}

\begin{itemize}
  \item $i=1$. Trivial, since $\ket {\Psi^\cF_1} = \ket {\Phi^\cF_1}$.

  \item $i>1$. Now, we assume that the Eq.\@~(\ref{eq:rec}) holds for
  $k<i$.  By definition of $N_i$ we have,
  \begin{eqnarray}
  N_i^2 & = & 1 - \sum_{k<i} \left| \bk{\Phi^\cF_i}{\Psi^\cF_k}
  \right|^2 \\ & \geq & 1 - (i-1) (\gamma^\cF)^2 + O((\gamma^\cF)^3).
  \end{eqnarray}

  Moreover, with Eq.\@~(\ref{eq:GS}) we also have,
  \begin{equation*}
  \left|\bk {\Phi^\cF_j}{\Psi^\cF_i}\right| = \frac{1}{N_i} \left|
  \bk{\Phi^\cF_j}{\Phi^\cF_i} - \sum_{k < i}\bk
  {\Psi^\cF_k}{\Phi^\cF_i} \bk{\Phi^\cF_j}{\Psi^\cF_k}\right|.
  \end{equation*}

  \begin{enumerate}
    \item When $j<i$ we get
    \begin{eqnarray}
      \left|\bk {\Phi^\cF_j}{\Psi^\cF_i}\right| &\leq& \frac{1}{N_{i}}
      \left( \left|\bk{\Phi^\cF_j}{\Phi^\cF_i}\right| \phantom{\sum_i}
      \right. \nonumber \\ && + \, \left|\sum_{k<i, \ k\neq j}
      \bk{\Psi^\cF_k}{\Phi^\cF_i}\bk{\Phi^\cF_j}{\Psi^\cF_k}\right|
      \nonumber \\ && \left. +\,
      \left|\bk{\Psi^\cF_j}{\Phi^\cF_i}\bk{\Phi^\cF_j}{\Psi^\cF_j}\right|
      \phantom{\sum_i}\right) \\ &\leq& 2 \gamma^\cF +
      O((\gamma^\cF)^2).
    \end{eqnarray}

  \item For $i=j$,
  \begin{eqnarray}
    \left|\bk {\Phi^\cF_i}{\Psi^\cF_i}\right| & = & {N_i} \\ & \geq &
    1 - \frac{i-1}{2} (\gamma^\cF)^2 + O((\gamma^\cF)^3).
    \end{eqnarray}

  \item Finally, in the case $j>i$, the contribution at order
  $\gamma^\cF$ comes from $\bk{\Phi^\cF_j}{\Phi^\cF_i}$ which gives,
  \begin{equation}
    \left|\bk {\Phi^\cF_j}{\Psi^\cF_i}\right| \leq \gamma^\cF +
    O((\gamma^\cF)^{2}).
  \end{equation}
\end{enumerate}
\end{itemize}

With this result, we can easily calculate an upper bound on $\| \rho -
\sigma \|_2$:
\begin{eqnarray}
\Tr\, (\rho- \sigma)^2 & = & \Tr\, \rho^2 + \Tr\, \sigma^2 -2\Tr\,
\rho\sigma \\ & = & 2 \sum_i |\alpha_i|^4 (1 -
|\bk{\Phi^\cF_i}{\Psi^\cF_i}|^2) \nonumber \\ && + \sum_{i\neq j}
|\alpha_i|^2|\alpha_j|^2|\gamma^{\overline\cF}_{ij}|^2 \nonumber \\ & \leq & 2
(d^\cS-1) (\gamma^\cF)^2 + (\gamma^{\overline \cF})^2 +
O((\gamma^\cF)^3) \nonumber
\end{eqnarray}
Therefore we have $\| \rho- \sigma\|_2 \leq \sqrt{2(d^\cS-1)
(\gamma^\cF)^2 + (\gamma^{\overline \cF})^2}+ O((\gamma^\cF)^{3/2})$, concluding the proof.
\end{proof}

\end{document}